\documentclass[twocolumn,english,pra,showpacs,preprintnumbers]{revtex4-1}
\usepackage{color}
\usepackage{amsmath}
\usepackage{amssymb}
\usepackage{graphicx}
\usepackage{tikz}
\usepackage{url}

\makeatletter

\@ifundefined{textcolor}{}
{%
 \definecolor{BLACK}{gray}{0}
 \definecolor{WHITE}{gray}{1}
 \definecolor{RED}{rgb}{1,0,0}
 \definecolor{GREEN}{rgb}{0,1,0}
 \definecolor{BLUE}{rgb}{0,0,1}
 \definecolor{CYAN}{cmyk}{1,0,0,0}
 \definecolor{MAGENTA}{cmyk}{0,1,0,0}
 \definecolor{YELLOW}{cmyk}{0,0,1,0}
}

\usepackage{textcomp}
\usepackage{dcolumn}
\usepackage{bm}
\usepackage[right]{eurosym}
\usepackage{float}
\usepackage[english]{babel}
\usepackage{blindtext}

\makeatother

\usepackage{babel}
\begin{document}

\title{Quantum dynamics of Rb atoms desorbing off the surface of He nanodroplets}

\author{N. V. Dozmorov}
\affiliation{Institute of Chemical Kinetics and Combustion, Novosibirsk, Russia}
\affiliation{Novosibirsk State University, Russia}
\author{J. von Vangerow}
\affiliation{Physikalisches Institut, Universit{\"a}t Freiburg, Germany}
\author{J. A. M. Fordyce}
\affiliation{Department of Physics and Astronomy, The University of British Columbia, Vancouver, Canada}
\author{F. Stienkemeier}
\affiliation{Physikalisches Institut, Universit{\"a}t Freiburg, Germany}
\author{A. V. Baklanov}
\affiliation{Institute of Chemical Kinetics and Combustion, Novosibirsk, Russia}
\affiliation{Novosibirsk State University, Russia}
\author{M. Mudrich}
\email{mudrich@phys.au.dk}
\affiliation{Department of Physics and Astronomy, Aarhus University, Denmark}

\begin{abstract}
The desorption of excited rubidium (Rb) atoms off the surface of helium (He) nanodroplets is studied in detail using femtosecond time-resolved photoion and photoelectron imaging spectroscopy in combination with quantum wave packet simulations. The good agreement of the measured time-dependent velocity distributions with the simulation when exciting the Rb dopant atoms into the 6p-state supports the pseudo-diatomic model (PDM) for the Rb-He droplet interaction, even on the level of quantum wave packet dynamics. Time-resolved photoelectron spectra reveal the partitioning of excitation energy into the dopant and the droplet degrees of freedom. 
\end{abstract}


\date{\today}

\maketitle

\section{Introduction}
An important research direction in the field of quantum fluid clusters is the study of dynamical processes in pure and doped helium nanodroplets induced by femtosecond laser excitation~\cite{Stienkemeier:2006,Mudrich:2014,Ziemkiewicz:2015,Ancilotto:2017}. On the one hand, He nanodroplets offer the opportunity to probe the dynamical response of a superfluid to an impulsive perturbation, both theoretically and experimentally~\cite{Shepperson:2017}. On the other hand, rich chemical dynamics are initiated by electronically exciting an impurity atom or a molecule inside or at the surface of a He nanodroplet. In this respect, doped He nanodroplets can serve as model systems to study solvent-induced photodynamics under rather well controlled conditions. Relaxation pathways in these systems include transfer of population between electronic states~\cite{Loginov:2007,Fechner:2012,Kautsch:2013,Lindebner:2014}, dissociation and recombination~\cite{Braun:2004,Kautsch:2015}, as well as complex formation~\cite{Reho:2000,Bruehl:2001,Droppelmann:2004,Leino:2011,Goede:2013,Lackner:2018}. As a general trend, electronically excited dopant atoms and small molecules tend to be ejected out of He droplets due to Pauli repulsion of the electron in the more diffuse excited state of the dopant from the surrounding He~\cite{Stienkemeier:2006,Loginov:2007,Brauer:2013,Koch:2014,Mudrich:2014,Takayanagi:2017}.

Alkali metal atoms are particularly well suited for studying the dynamics of laser-induced desorption due to their well-defined initial position in dimple-like states at the surface of He nanodroplets~\cite{Stienkemeier:1996,Barranco:2006}. Upon electronic excitation, all alkali species promptly desorb off the droplet surface, with the only exceptions being rubidium (Rb) and cesium atoms in their lowest excited states~\cite{Auboeck:2008,Theisen:2011}. The dynamics of the desorption process have recently been studied at an increasing level of detail~\cite{Takayanagi:2004,Fechner:2012,Hernando:2012,Vangerow:2014}, including time-resolved experiments and simulations~\cite{Vangerow:2015,Vangerow:2017,Coppens:2018}. Many of the spectral and dynamical aspects of this process have been surprisingly well reproduced by the simple pseudo-diatomic model (PDM). In this model, the dopant-He droplet complex is represented by a diatomic molecule, where the entire He droplet is treated as a single constituent atom~\cite{Stienkemeier:1996,Reho:2000,LoginovPRL:2011,Callegari:2011}. Clearly, the PDM neglects important aspects of the dynamics such as local deformations of the He density distribution in the vicinity of the excited dopant, as well as electronic relaxation. Nevertheless, for Rb atoms attached to He nanodroplets, we were able to reproduce the general trends of the experimental velocity and time-resolved photoionization signals using the PDM~\cite{Fechner:2012,Vangerow:2015}. In particular, the interplay between desorption of the excited Rb dopant atom and the falling back of the Rb$^+$ and [RbHe]$^+$ ions into the He droplet when ionization occurs at short pump-probe delay times was unambiguously established by comparing to classical trajectory simulations based on the PDM~\cite{Vangerow:2015}. However, these simulations yielded systematically shorter time constants for the desorption dynamics. Time-dependent density functional simulations essentially supported our concept of the fall-back effect~\cite{Vangerow:2017,Coppens:2018}. For the lowest excited state of Rb, 5p, deviations of the observed signals from the simulated dynamics were found, indicating that for weakly repulsive states, excited atoms are ejected by an evaporation-like process rather than by impulsive desorption according to the PDM~\cite{Vangerow:2017}. The desorption of RbHe exciplexes was found to be induced by He droplet-mediated spin-relaxation~\cite{Coppens:2018}. 

The aim of this work is to take the study of the desorption dynamics of excited Rb atoms to the next level of detail, both experimentally and theoretically. To this end, we analyze the velocity and angular distributions of Rb$^+$ and [RbHe]$^+$ ions as well as photoelectrons for the previously studied Rb excited states, and we compare them to quantum wave packet simulations. In this way, we scrutinize the role of quantum dynamics in this nanometer-scale system, as well as the validity of the PDM in this approach.

\section{Methods}
\label{sec:Methods}
The setup used for the present experiments has been described previously~\cite{Vangerow:2014,Vangerow:2015}. Briefly, a beam of He droplets with an average diameter of 10~nm is produced by continuously expanding pressurized He (50~bar) out of a cold nozzle (diameter 5~$\mu m$, temperature 17~K). The He droplets are doped with one Rb atom on average by pickup of Rb atoms inside a heated vapor cell (length 1~cm, temperature 85$^\circ$C). 

An amplified Ti:Sa laser system generates pulses of about 0.4~mJ pulse energy at a repetition rate of 5~kHz. The FWHM pulse duration is 120 fs with a variation of 20 fs for different laser center wavelengths. The center wavelength is tunable in the near infrared (NIR) region. Light in the visible range (VIS) is generated by frequency doubling. The pulses are split and separated in time in a Mach-Zehnder type interferometer. The FWHM of the cross correlation signals of the two VIS pulses and of one NIR and one VIS pulse is about 150~fs. The droplet-perturbed Rb 5p and 6p states are probed by a two-color NIR+VIS and a one-color VIS resonant pump-probe photoionization scheme, respectively. In the NIR+VIS scheme, the NIR pulses are strongly attenuated to avoid the excitation of higher lying states by multi-photon processes. 

Photoions and photoelectrons are detected by a velocity-map imaging (VMI) spectrometer operated in single-particle detection mode~\cite{Eppink:1997,Fechner:2012,Vangerow:2014,Coppens:2018}. Mass-selected ion VMIs are recorded for varying pump-probe delay steps. The Rb$^+$ ion VMIs contain a background contribution from the near-resonant ionization of free Rb atom effusing out of the doping cell. In the measurements of the Rb 6p states, additional background signal arises from the ionization of free and droplet-bound Rb by the single laser pulses. To extract the pump-probe correlated contribution to the total signal, the mentioned background contributions are subtracted from each recorded image. The photoelectron spectra obtained from photoelectron VMI are treated in the same way.

\begin{figure}
	\centering \includegraphics[width=0.48\textwidth]{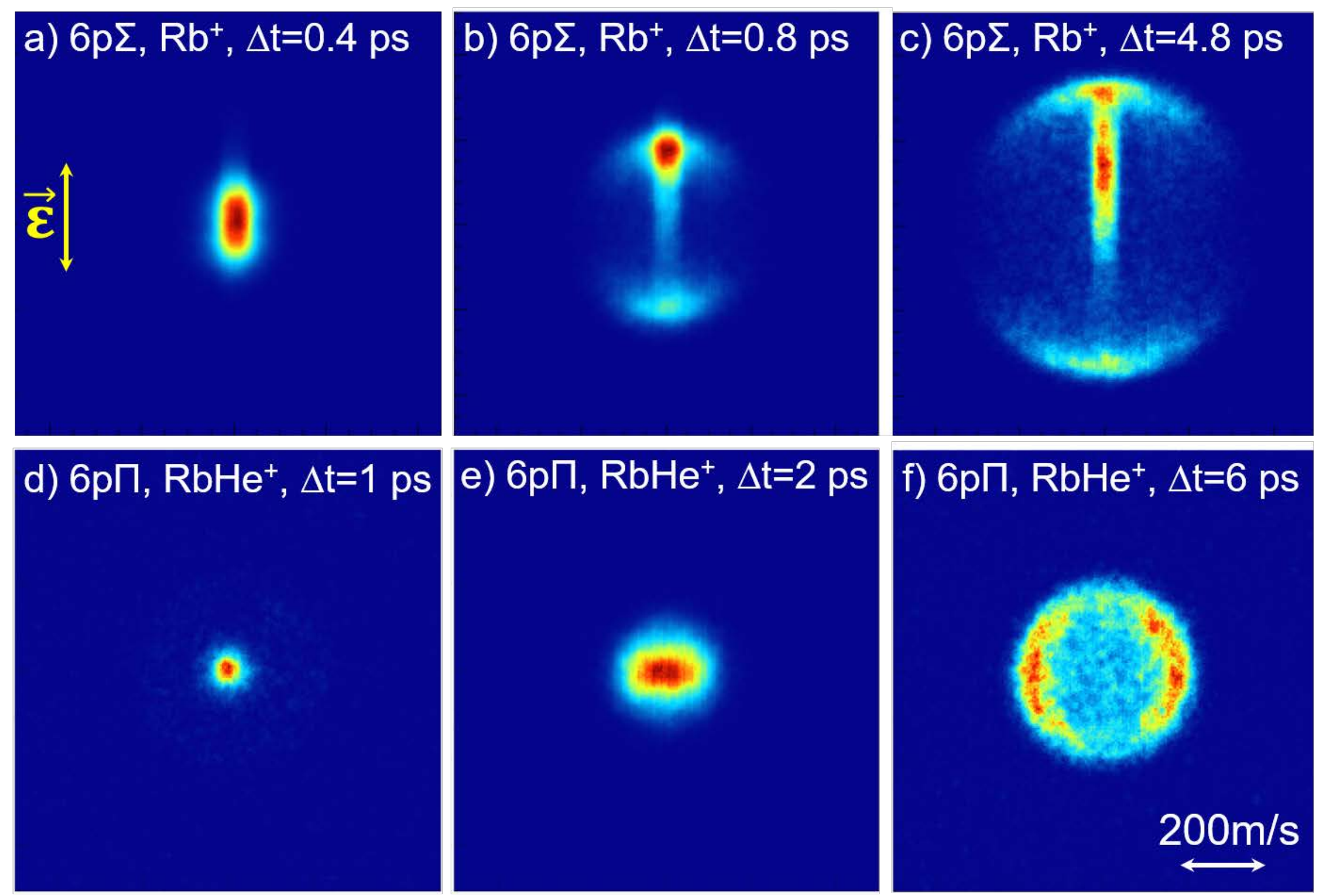}
	\protect\caption{\label{fig:VMIs} Velocity map images of Rb$^+$ ions [a)-c), $\lambda = 403~$nm (6p$\Sigma_{1/2}$-state)] and of [RbHe]$^+$ ions [d)-f), $\lambda = 415~$nm (6p$\Pi$-state)], recorded at different delays between pump (excitation) and probe (ionization) pulses, $\Delta t$. The vertical double-sided arrow in a) depicts the polarization direction of the laser light.}
\end{figure}
To infer speed distributions and anisotropy parameters, the raw VMIs are inverse Abel transformed using the maximum entropy routine MEVELER developed by B. Dick~\cite{Dick:2014}. In the case of detecting Rb$^+$, only the lower half of each VMI (see Fig.~\ref{fig:VMIs}) is used because in this region the two signal components (effusive, desorbed off He droplets) are well separated. When detecting [RbHe]$^+$ ions we use the complete VMIs of [RbHe]$^+$ in our analysis procedure because there is no [RbHe]$^+$ signal contribution from effusive Rb. The speed distributions are then fitted by skewed gaussian functions to infer the relevant characteristics of the distributions~\cite{mudholkar2000epsilon}.

The wave packet dynamics of the excited and ionized Rb-He droplet system is simulated by solving the one-dimensional Schr{\"o}dinger equation for pseudo-diatomic Rb-He droplet interaction potentials using the split-operator method~\cite{Feit:1982}. For that, the Wavepacket Matlab library realization of the split operator method is employed~\cite{Schmidt:2017}. This library has been modified to account for the interaction of wave packets with a laser field within perturbation theory~\cite{TannorBook}. For the 6p-states, a uniform grid from 0 to 800~\AA{} with a spacing of 2.4~pm is used and the time step is 0.1~fs. For the 5p-states, a uniform grid from 0 to 1600~\AA{} with the same spacing is used and the time step is 0.05~fs. For each value of the pump-probe delay, the full propagation is carried out until convergence of the velocity distribution is reached (60~ps - 1~ns). 

\begin{figure}
	\centering \includegraphics[width=0.45\textwidth]{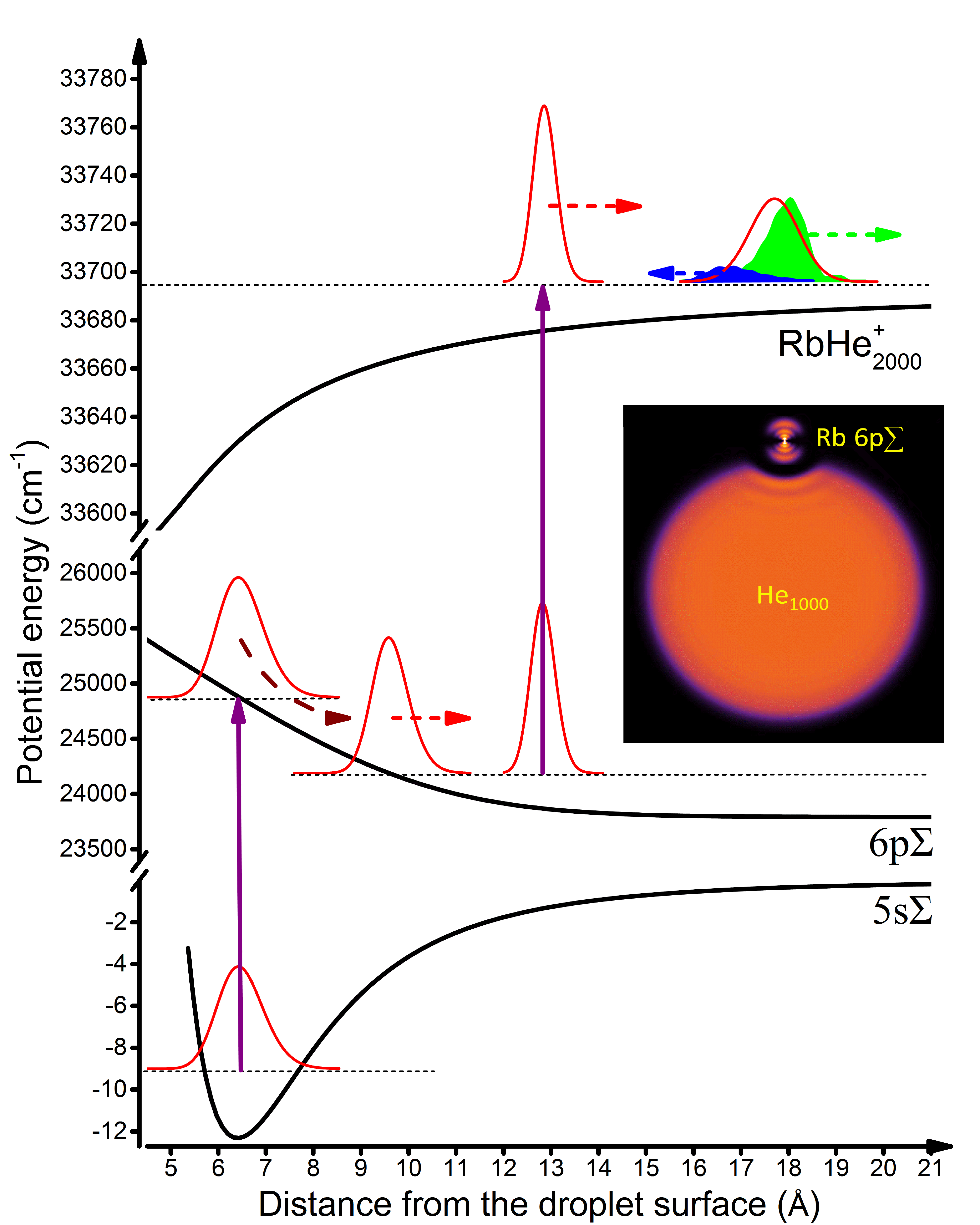}
	\protect\caption{\label{fig:potscheme}
		Pseudodiatomic potential energy curves involved in the desorption process. The red bell-shaped curves depict the wave functions as they propagate in time. The inset illustrates the configuration of a Rb atom excited to the 6p$\Sigma_{1/2}$-state at the surface of a He nanodroplet containing 1000 atoms; adapted from \cite{Coppens:2018}.}
\end{figure}
The potential energy curves (PECs) of the neutral Rb-He droplet complex are taken from Ref.~\cite{Callegari:2011}, where spin-orbit splitting of the excited state is included using the procedure outlined in Ref.~\cite{Reho2:2000}. The PEC of the Rb$^+$-He droplet complex is obtained from the Rb$^+$-He pair potential~\cite{Koutselos:1990} on the basis of the He density distribution of the groundstate RbHe$_{2000}$ complex~\cite{Pi}. Thus, we consider all PECs to be constant during the propagation of the wave packet. In particular, transient modulations of the He droplet surface are neglected. Fig.~\ref{fig:potscheme} displays these PECs, including snapshots of the wave packet propagation in the excited 6p$\Sigma_{1/2}$-state and in the ionic state of the RbHe$_{2000}$ complex. 

The initial wave function of the excited Rb-He droplet complex is obtained from the speed distribution of Rb$^+$ ions measured at long delay time ($\Delta t=10$~ps for the measurements of the 6p-state, $\Delta t=250$~ps for the 5p-state) where it converges to a stationary distribution. Due to the conservation of energy in the wave packet propagation, this final speed distribution relates to an initial spatial distribution $\left|\psi_*(R)\right|^2$ of the wave packet on the repulsive PEC at short distances. This approach differs from the one we used previously for a mixed quantum-classical simulation~\cite{Vangerow:2015}. There, the wave packet propagation was initialized based on the spectral profile of the pump laser pulse. Compared to that approach, in the present simulations the initial wave packet is located at slightly larger distance of the Rb atom from the He droplet surface $R$ where the potential energy is lower. Upon excitation into the most repulsive 6p$\Sigma_{1/2}$-state at a laser wavelength $\lambda=403$~nm, the initial position is $R=9.4$~\AA{} as opposed to $R=6.4$~\AA{} in the previous approach. This fast shifting of the Rb-He droplet distance is represented by the curved, dashed arrow in Fig.~\ref{fig:potscheme}. The present model assumptions can be rationalized by a fast initial rearrangement of the He droplet surface in the vicinity of the excited Rb atom. Due to the strong $R$-dependence of the Rb-He interaction, the Rb atom is expected to interact predominantly with only a few neighboring He atoms in the moment of excitation~\cite{Vangerow:2014,Vangerow:2015}. Following this initial phase of local compression of the He droplet surface, the Rb atom interacts with the He droplet as a whole. Therefore, in the wave packet propagation, the reduced mass of the system is taken as that of one Rb atom interacting with 2000 He atoms, which is close to the mass of one Rb atom.

$\psi_*(R)$ is calculated as a superposition of vibrational eigenstates $\phi_i(R)$ in the excited state PEC, 
\begin{equation*}
\psi_*(R)=\sum_{i=0}^{N} (-1)^i \sqrt{a(E_i)} \phi_i(R).
\end{equation*}
To obtain discrete eigenstates $\phi_i(R)$, the excited state PEC is augmented by an attractive branch at sufficiently large distance such that the dynamics remains unaffected. $\phi_i(R)$ are calculated numerically using the program LEVEL 8.0 ~\cite{level}. The coefficients $a(E_i)$ are given by the Rb kinetic energy spectrum inferred from the experimentally determined final speed distribution. The propagation of this wave packet is performed in the unchanged PEC.

In the transition to the ionic state induced by the probe pulse, the kinetic energy distribution is conserved due to the Franck-Condon principle. Subsequently, this wave packet propagates in the ionic state where it splits into two partial wave packets. One of the two corresponds to Rb$^+$ ions that escape from the He droplet surface, and the other corresponds to Rb$^+$ that falls back into the He droplet interior. The splitting of the ionic state wave packet in two is illustrated in Fig.~\ref{fig:potscheme} (upper right) by the blue and the green filled distributions. Thus, the wave packet in the ionic state $\psi_+(R)$ is represented as a superposition of discrete eigenstates $\phi_+$ in the bound part of the ionic PEC, and of continuum eigenstates $\phi_{E}$ in the energy region above the dissociation limit,
\begin{equation*}
\psi_+ = \sum_{i=0}^{N} c_i \phi_i^+ + \int_{0}^{\infty} c(E) \phi_{E} dE.
\end{equation*}   

From the bound part of the wave packet we calculate the fraction of Rb$^+$ ions that escape the He droplet as $N_{Rb^+, \,\mathrm{escape}} = 1 - \sum_{i=0}^{N} c_i^2$, where $c_i = \big \langle\phi_i^+ | \psi_+\big \rangle$. Accordingly, the wave function of ions that escape from the droplet is given by $\psi_{Rb^+,\,\mathrm{escape}}=\psi_+ -\sum_{i=0}^{N} c_i \phi_i^+$. The speed distributions, which we analyze in terms of position, width and skewness, are obtained from $\psi_{Rb^+,\,\mathrm{escape}}$ using fast Fourier transformation. All numerical calculations besides the wave packet propagation are performed using Wolfram Mathematica~\cite{Mathematica}. 

The simulated photoelectron distributions $p(E)$ are obtained by a convolution to account for the spatial extension of the wave packet and the spectral profile of the ionizing laser pulse,
\begin{equation*}
 p(E) = \int_0^{\infty} |\psi_*(R)|^2 exp(-\frac{(h\nu - U_+(R) + U_*(R) - E)^2}{2\sigma_\mathrm{las}^2}) dR.
\end{equation*} 
Here, $h\nu$ and $\sigma_\mathrm{las}=140$~cm$^{-1}$ are the center photon energy and the spectral width of the probe pulse, respectively, and $U_*(R)$ and $U_+(R)$ denote the excited and ionic state PECs, respectively.

\section{Results and discussion}
By measuring VMIs of photoions and photoelectrons created by the probe laser pulse as a function of delay $\Delta t$ with respect to the pump pulse, detailed insight into the dynamics of the desorption process is obtained. Fig.~\ref{fig:VMIs} shows examples of VMIs of Rb$^+$ ions [a)-c)] and [RbHe]$^+$ ions [d)-f)] for selected values of $\Delta t$. The laser polarization axis is indicated by the arrow in a). The direction of propagation of the droplet beam is also vertical, from top to bottom. The center wavelength of both excitation and ionization laser pulses is $\lambda$=403~nm in [a)-c)], which leads to the excitation of the most repulsive 6p$\Sigma_{1/2}$ pseudodiatomic state. In [d)-f)], the wavelength is $\lambda$=415~nm which corresponds to the 6p$\Pi$-state, where efficient formation of RbHe exciplexes has been found~\cite{Fechner:2012,Vangerow:2014,Vangerow:2015}.  

The Rb$^+$ VMIs in Fig.~\ref{fig:VMIs} a)-c) feature a ring-like structure due to Rb atoms that have desorbed off the surface of He droplets and have gained a well-defined velocity in the moving frame of the droplet beam. The vertical streak-like distribution stems from direct off-resonant two-photon ionization of Rb atoms which are effusively emitted from the doping cell and propagate along the droplet beam axis. 

The angular distribution of the signal intensity is clearly different for the Rb$^+$ and [RbHe]$^+$ signals. In the Rb$^+$ VMIs ($\lambda = 403~$nm) the signal is peaked along the polarization direction of the laser. This is due to the fact that desorption occurs in the 6p$\Sigma_{1/2}$-state of the Rb-He droplet complex. Since the initial state (5s$\Sigma_{1/2}$) has the same symmetry, $\Sigma_{1/2}$, a $\cos^2\theta$ angular dependence (anisotropy parameter $\beta=2$) is expected if the PDM were perfectly valid~\cite{Zare:1972,Fechner:2012}. Here, $\theta$ is the angle between the laser polarization and the ion velocity in the detector plane. The [RbHe]$^+$ images recorded at $\lambda = 415~$nm reflect the desorption dynamics in the 6p$\Pi$-state, for which we expect a $\sin^2\theta$-dependence ($\beta = -1$)~\cite{Fechner:2012}. At intermediate wavelengths, superpositions of $\Sigma$ and $\Pi$-states are excited, and the ion VMIs are closer to isotropic ($\beta = 0$).

\begin{figure}
	\centering \includegraphics[width=0.45\textwidth]{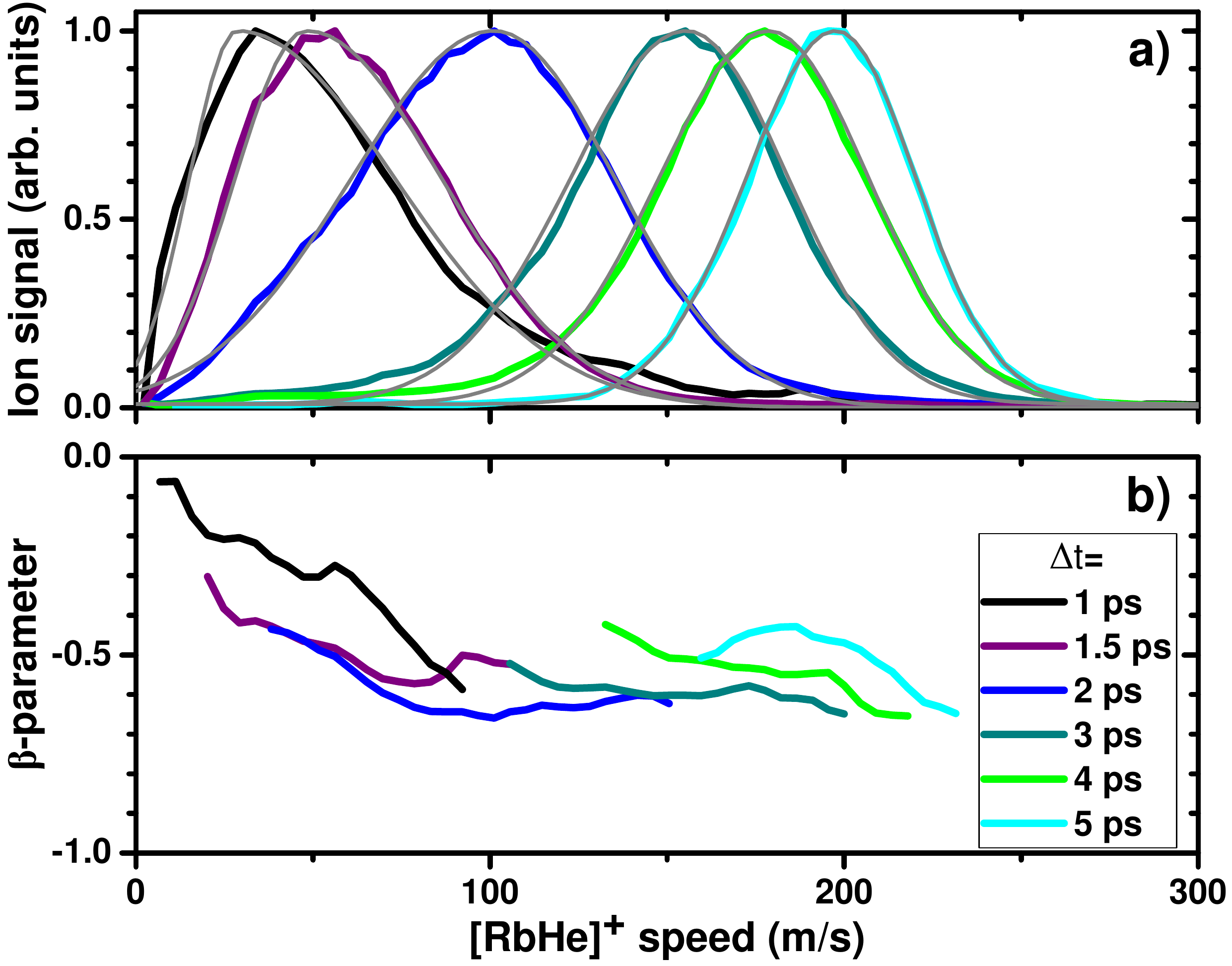}
	\protect\caption{\label{fig:speed}
	Speed distributions (a) and anisotropy parameters $\beta$ (b) of [RbHe]$^+$ ions recorded at different pump-probe delays. The excitation wavelength is centered at $\lambda = 415$~nm, corresponding to the 6p$\Pi$ state. Gray lines indicate skewed Gaussian fit functions at different delays $\Delta t$.}
\end{figure}
The speed distributions and $\beta$-parameters infered from the [RbHe]$^+$ ion VMIs recorded at $\lambda = 415$~nm for various values of $\Delta t$ are depicted in Fig.~\ref{fig:speed} a) and b), respectively. The thin grey lines in a) are best fits to the data using skewed Gaussian functions as a model~\cite{Mudholkar:2000}. As $\Delta t$ increases from 1 to 6~ps, the speed distribution shifts from around 50 to 200~m/s. The rising of the recorded ion speeds as a function of $\Delta t$ was previously discussed for Rb$^+$ ions in the context of the competing desorption of the excited Rb atom and the falling back of the photoion into the droplet~\cite{Vangerow:2017}. Besides, we clearly see a change in the shape of the speed distribution from highly asymmetric to symmetric with a gaussian-like shape. In that range of $\Delta t$, the $\beta$-parameters evolve from values near zero (isotropic) to values around $-0.5$.

\begin{figure}
	\centering\includegraphics[width=0.47\textwidth]{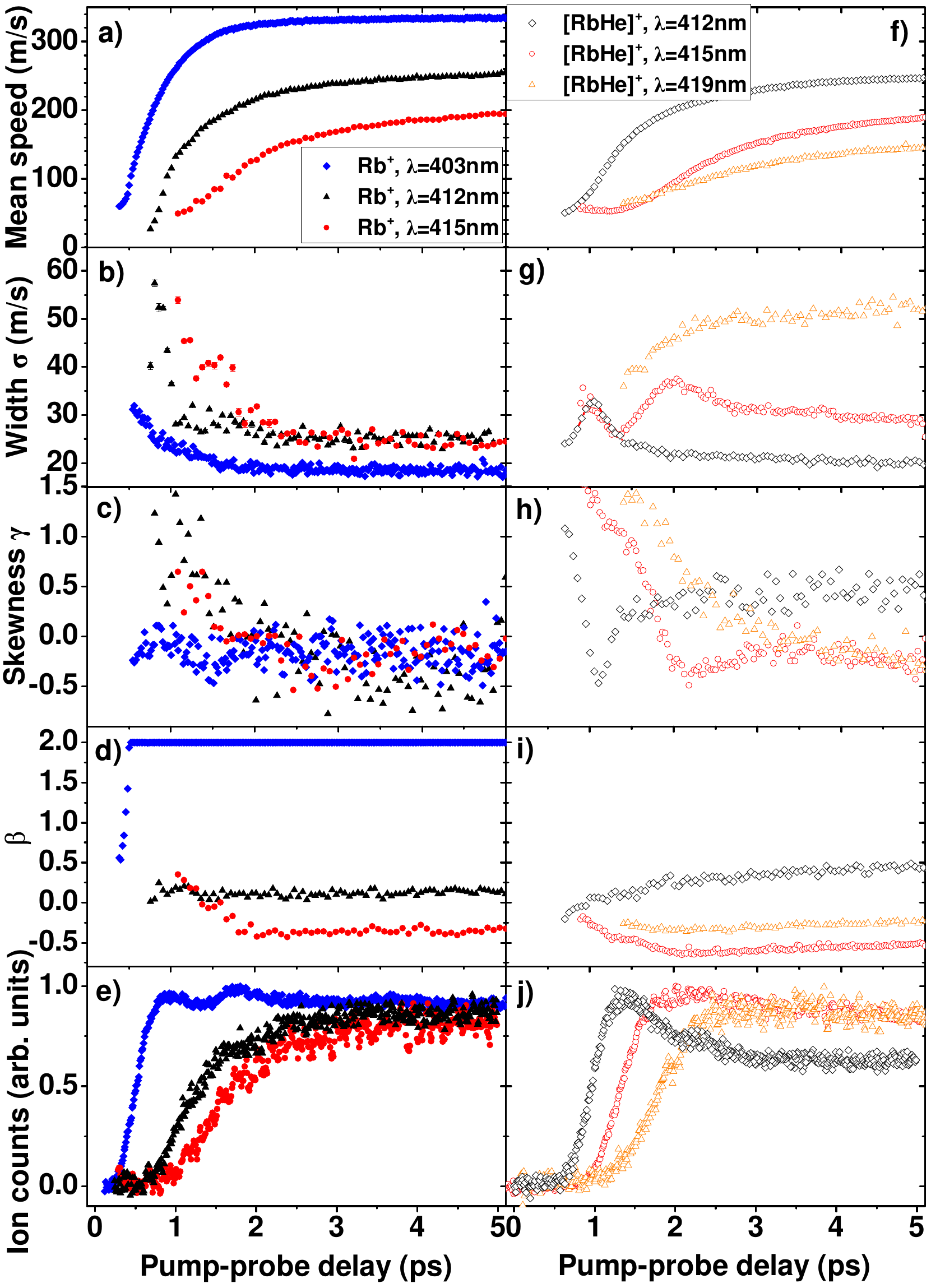}
	\protect\caption{\label{fig:exp_ions}Overview over the experimental parameters derived from the Rb$^+$ and [RbHe]$^+$ delay-dependent ion images. a) Mean speed, b) standard deviation $\sigma$ of speed distribution, c) skewness $\gamma$ of speed distribution, d) anisotropy parameter $\beta$, and e) ion yields. The center wavelengths of the laser are given in the legends.} 
\end{figure}
Owing to the high quality of our imaging data, this analysis procedure allows us to infer from the speed distributions not only the average value of the speed, but also second and third statistical moments; that is the standard deviation $\sigma$ and the skewness $\gamma$. An overview of these data measured by detecting Rb$^+$ ions (left column) and [RbHe]$^+$ ions (right column) for three different wavelengths of the laser $\lambda$ is presented in Fig.~\ref{fig:exp_ions}. The evolution of the yields of photoions [panels e) and j)] have been discussed previously and are shown here for completeness~\cite{Vangerow:2015}. The delayed increase of ion yields in the range $0.4\leq \Delta t \leq 1.6$~ps is due to the falling back of ions into the He droplet when ionization by the probe pulse occurs while the excited, desorbing Rb atom or RbHe exciplex still is in the vicinity of the He droplet. Since the interaction of ions with the He droplet is strongly attractive, as opposed to the interaction of excited neutrals with the He droplet surface which is generally repulsive, ions created by the probe pulse at short delays tend to fall back into the droplet thereby reducing their count rates. The delay corresponding to the transition from predominant fall-back to the ejection of free ions is controlled by the laser wavelength. The latter determines (i) which electronic state of the RbHe$_N$ complex is excited, and (ii) at which Rb-He droplet distance for a specific electronic state the desorption dynamics is initiated. Therefore, upon excitation at the shortest wavelength $\lambda=403$~nm (blue symbols), the fall-back time is shortest (0.4~ps), and at $\lambda=419$~nm (orange symbols) the fall-back time is longest (1.9~ps). At $\lambda=403$~nm, the most repulsive 6p$\Sigma_{1/2}$-state is excited, which can be seen from the anisotropy parameter $\beta=2$ shown in Fig.~\ref{fig:exp_ions} d). For $\lambda=412$, 415, and 419~nm, $\beta$ ranges between $-0.5$ and $0.5$, due to the excitation of a superposition of 6p$\Sigma_{1/2}$ and 6p$\Pi$-states at these wavelengths. 

As the ion yields increase with delay time, the mean speeds of ions also rise as a result of competing attractive and repulsive dopant-He droplet interactions [Figs.~\ref{fig:exp_ions} a) and f)]~\cite{Vangerow:2017}. The slower rise of ion speeds compared to ion yields is due to the fact that even at Rb-He droplet distances beyond the point where fall-back occurs, ion-He attractive forces continue to be active and slow down the motion of the free ion. The fact that the [RbHe]$^+$ ion speeds and yields closely follow those of Rb$^+$ ions indicates that RbHe exciplex formation occurs on a much shorter time scale than the desorption dynamics~\cite{Fechner:2012, Vangerow:2015}. Thus, given the small difference in mass between Rb and RbHe, both the desorption dynamics of excited Rb and RbHe, and the fall-back of ionized Rb$^+$ and [RbHe]$^+$ proceeds essentially in the same way. This is in contrast to the 5p excited state, where qualitatively different dynamics for Rb and RbHe were observed~\cite{Coppens:2018}. 

More detailed insights into this dynamics is obtained from the width $\sigma$ and the skewness $\gamma$ of the ion speed distributions shown in panels b), c), g) and h) of Fig.~\ref{fig:exp_ions}. Since the systematic uncertainty is lower for the [RbHe]$^+$ data due to the absence of the effusive background, we will concentrate our discussion on the [RbHe]$^+$ data, assuming that the desorption dynamics of RbHe is essentially identical to that of Rb. Interestingly, both $\sigma$ and $\gamma$ show a more complex evolution as a function of delay time $\Delta t$ than the mean speed. While $\sigma$ features maxima at around $\Delta t=1$, $2$, and possibly $5$~ps for $\lambda=412$, $415$, and $419$~nm, respectively, $\gamma$ shows minima around the same values of $\Delta t$. This behavior cannot be explained in simple terms considering the classical motion of the excited and ionized Rb and RbHe dopants subjected to repulsive and attractive forces. Presumably, the shape of the speed distribution reflects the evolution of a quantum wave packet created by the pump pulse, as it is commonly assumed for diatomic molecules~\cite{Tannor:1985,Zewail:1994,Schmidt:2014}. 

\begin{figure}
	\centering \includegraphics[width=0.47\textwidth]{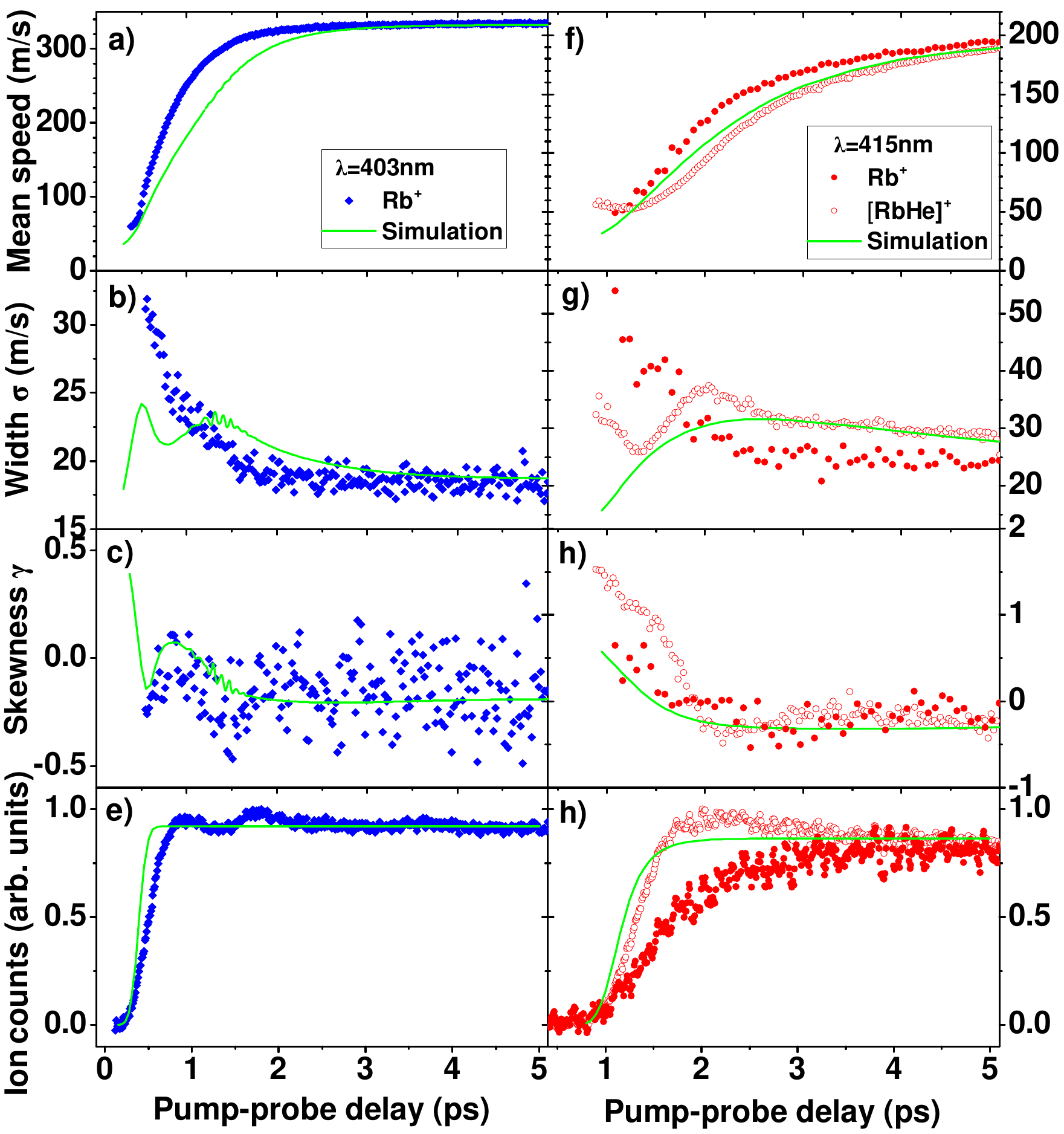}
	\protect\caption{\label{fig:transients6p}
		Comparison between experimental and simulated characteristics of the wave packet propagation of the excited and ionized RbHe$_N$ complex. At the laser wavelength $\lambda=403$~nm (left column), mainly the 6p$\Sigma_{1/2}$-state is excited; at $\lambda=415$~nm (right column), the 6p$\Pi$-state is excited and Rb$^+$ as well as [RbHe]$^+$ ions are detected.}
\end{figure}
To assess this conjecture, we perform numerical simulations of the wave packet propagation as outlined in Sec.~\ref{sec:Methods}. Fig.~\ref{fig:transients6p} shows a direct comparison between the characteristics of the experimental speed distributions and the results from the simulation for $\lambda = 403$~nm (left column) and for $\lambda = 415$~nm (right column). We find reasonably good agreement for all the displayed quantities. In particular, for  $\lambda = 415$~nm, which corresponds to excitation of mainly the 6p$\Pi_{3/2}$-state, the simulated rising edges of both the mean speed and the [RbHe]$^+$ ion yield match the experimental results very well [panels e) and h)]. The better agreement of experiment and simulation as compared to our previous model calculations~\cite{Vangerow:2015} is mainly due to the different initialization procedure of the excited wave packet, which supports the model assumptions made in the present approach. Furthermore, the simulated widths $\sigma$ and skewnesses $\gamma$ match the experimental values quite well for long delays $\Delta t$. At short $\Delta t$, we observe a drop of $\sigma$ as a function of $\Delta t$ for Rb$^+$ and a maximum around $\Delta t=2$~ps for [RbHe]$^+$. 

The simulated width $\sigma$ features two opposing trends that cause a maximum to appear around $\Delta t = 1.3$ and 2.5~ps for $\lambda=403$ and 415~nm, respectively.
The rise of the simulated width $\sigma$ at short $\Delta t$ up to the maximum reflects the increasing energy bandwidth of the truncated wave packet in the ionic state that spills over the threshold for escaping the droplet as a free ion. The trend of decreasing $\sigma$ for long $\Delta t$ is due to the increasing dispersion of the excited state wave packet in real space as it propagates along the short-range repulsive excited state PEC. Similarly, $\gamma$ features a maximum around $\Delta t=0.8$~ps for $\lambda = 403$~nm. The final values are slightly negative for both $\lambda = 403$ and $415$~nm. 


\begin{figure}
	\centering \includegraphics[width=0.35\textwidth]{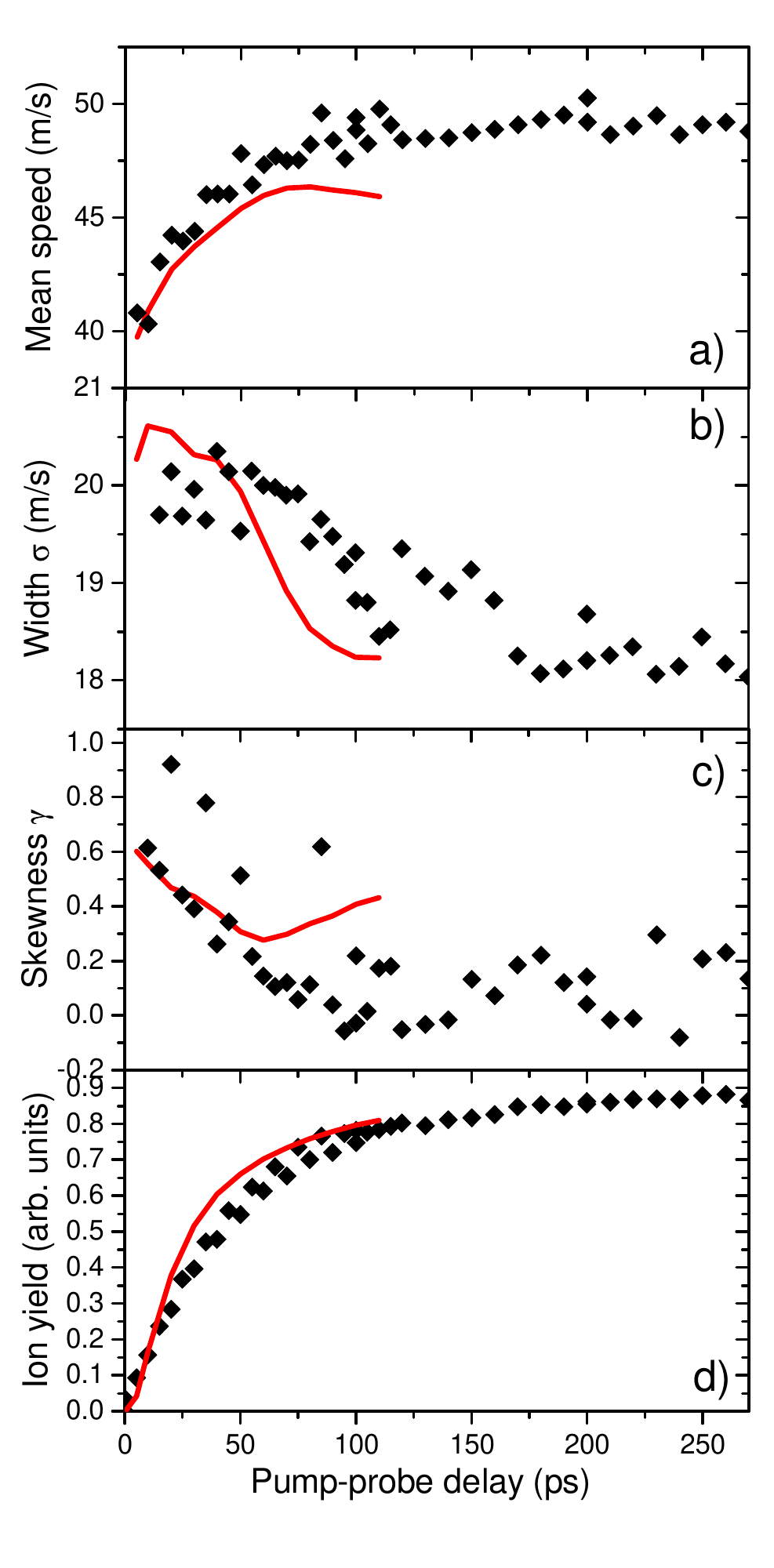}
	\protect\caption{\label{fig:transients5p}
		Comparison between experimental data for Rb$^+$ and the simulated wave packet characteristics for the 5p$\Sigma_{1/2}$-state of the RbHe$_N$ complex. The center wavelength of the laser is set to $\lambda = 773~$nm.}
\end{figure}
To test the general applicability of our simulation method we compare the experimental and simulation results for the lower 5p$\Sigma_{1/2}$-state of the Rb-He droplet complex which correlates with the atomic 5p$_{3/2}$-state. Although this state is the most repulsive of the triplet of 5p-correlated states, desorption proceeds nearly two orders of magnitude more slowly than in the 6p-correlated states~\cite{Vangerow:2017,Coppens:2018}. Nevertheless, we find good agreement between the experimental observables for Rb$^+$ ions and the results of the simulation in the range of delay times $0\lesssim\Delta t\lesssim 100$~ps. For longer simulation times, the accumulation of numerical errors make the simulation results unreliable. For the 5p$\Pi_{3/2}$-state (not shown), we find larger deviations, which we attribute to the fact that desorption does not proceed impulsively but is driven by spin-relaxation, which is not included in the present simulation~\cite{Coppens:2018}.

Besides speed distributions, the VMIs provide information about the angular distribution of ions and electrons. In the following, we address the measured angular anisotropy of the Rb$^+$ and [RbHe]$^+$ ions as a function of $\Delta t$, Fig.~\ref{fig:exp_ions} d) and i), respectively. Since at $\lambda=403$~nm the 6p$\Sigma_{1/2}$-state is excited with high purity, the anisotropy parameter takes the value $\beta=2$ over the entire range of $\Delta t$ where significant yields of Rb$^+$ ions are detected. At $\lambda=412$-$419$~nm, the asymptotic values of $\beta$ range between $-0.6$ and $0.5$ which can be rationalized by the fact that superpositions of 6p$\Sigma_{1/2}$ and 6p$\Pi$ are excited at these wavelengths. However, for $\Delta t\lesssim 2$~ps, we observe a drop of $\left|\beta\right|$ to zero near the onset of detected ion signals. 

In photodissociation of diatomic molecules, the anisotropy of recoil directions of the photofragments is governed by the angular anisotropy of the photon absorption, the lifetime of the excited molecule, the frequency of molecular rotation and the kinetic energy of the relative motion of the photofragments which fly apart~\cite{Jonah:1971,Yang:1974}. A reduction of $\beta$ is observed when either the molecule is highly rotationally excited, or when dissociation proceeds slowly in weakly repulsive states. In either of these cases, the anisotropy due to the radially flying apart of the dissociation products is blurred by the angular motion associated with molecular rotation. In the present experiment, the speed of the escaping Rb$^+$ ion is lowest at short $\Delta t$, which makes the blurring effect of the rotation of the Rb-He droplet complex on the angular anisotropy most pronounced. 

\begin{figure}
	\centering \includegraphics[width=0.45\textwidth]{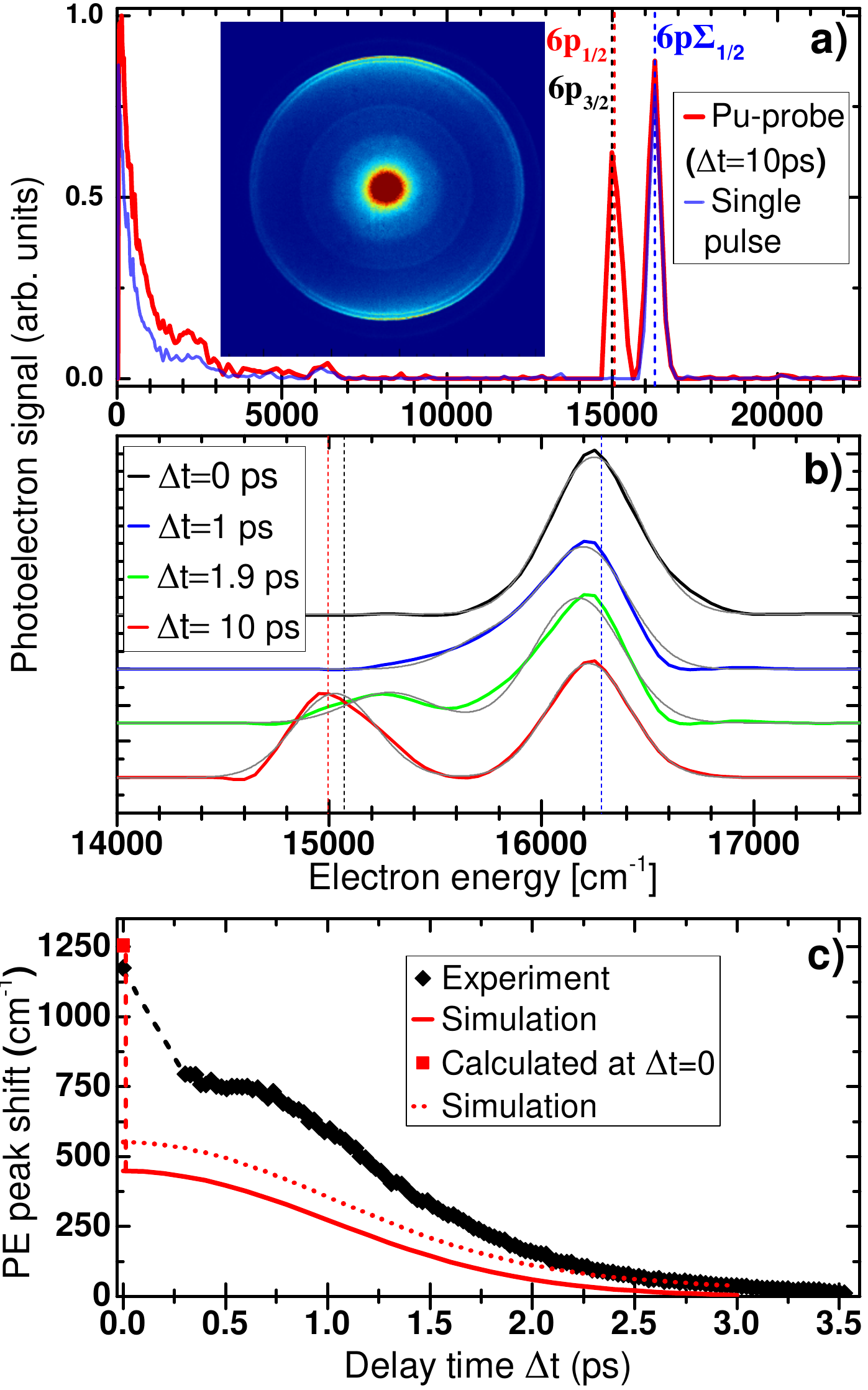}
	\protect\caption{\label{fig:PES}
		a) Photoelectron spectra recorded at $\lambda=401$~nm using single laser pulses (blue line) and pairs of pulses delayed by $\Delta t=10$~ps (red line). The vertical dashed lines indicate the 6p$_{1/2,\,3/2}$ atomic photolines (red and black) and the photoelectron energy for direct two-photon ionization (blue). The inset shows the corresponding electron VMI for two-pulse ionization. b) Close-up of the electron peaks at selected delays $\Delta t$. The spectra are vertically shifted for the sake of clarity. The gray lines depict fit curves. c) Experimental and simulated positions of the electron peak associated with the desorbing Rb and RbHe, referenced to the 6p$_{3/2}$-atomic state. }
\end{figure}
Complementary information about the relaxation dynamics of the excited Rb-He droplet complex is obtained from time-resolved photoelectron spectra. Such spectra are inferred from electron VMI recorded at the same experimental conditions as the ion VMI. A typical electron VMI recorded at $\lambda=401$~nm for a pump-probe delay $\Delta t=10$~ps is displayed in the inset of Fig.~\ref{fig:PES} a). The most prominent features are a bright central spot and two sharp outer rings. The central spot is due to electrons with very low kinetic energy. Such electrons have previously been observed in experiments where He droplets were directly excited or ionized~\cite{Peterka:2003,Buchta:2013,BuchtaJCP:2013}. The occurrence of slow electrons in this experiment points at photoelectrons undergoing massive scattering with the He droplet and possibly being transiently trapped in bubble states~\cite{Rosenblit:2006}. 

\begin{figure}
	\centering \includegraphics[width=0.5\textwidth]{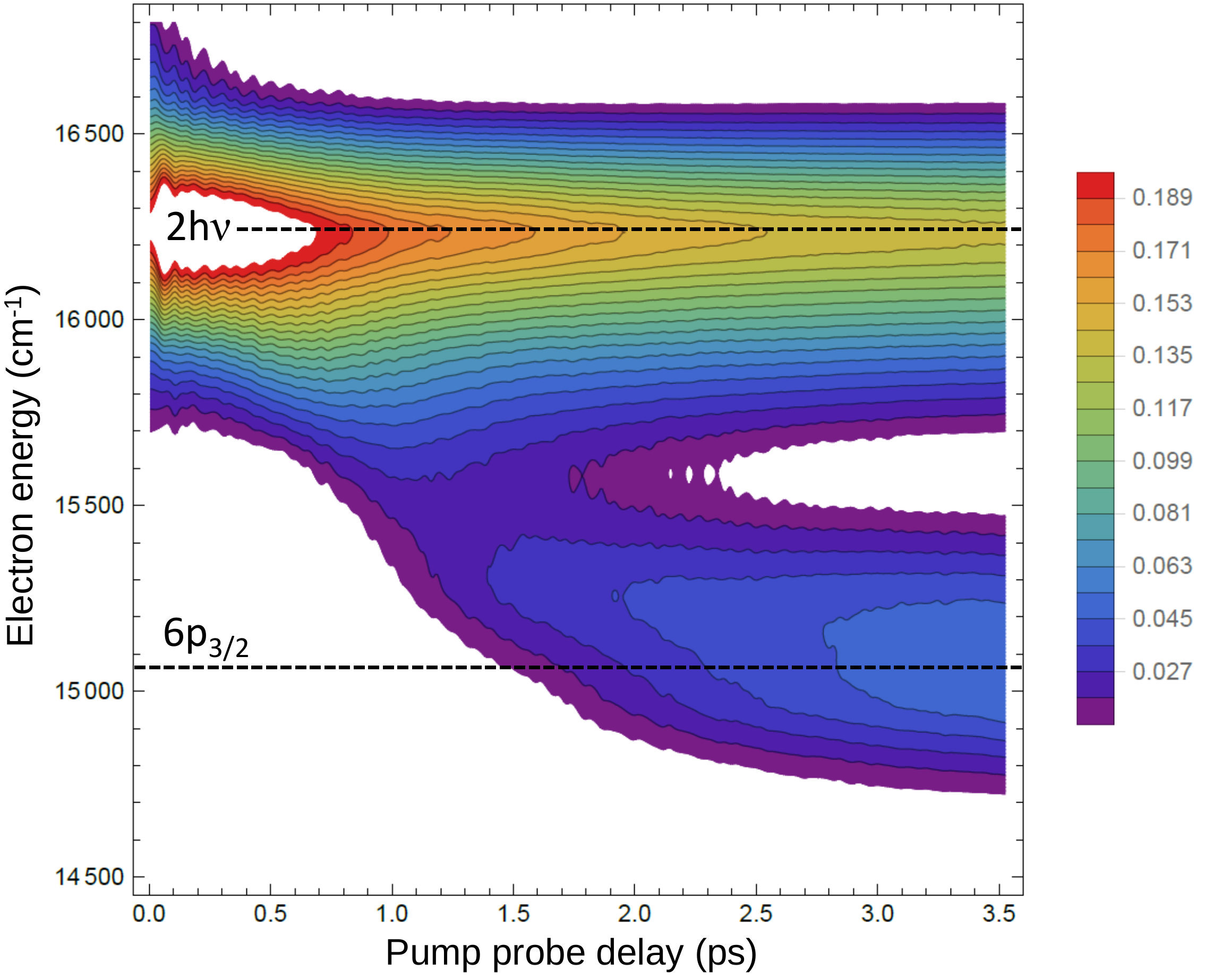}
	\protect\caption{\label{fig:Contour}
		Contour plot generated from the photoelectron spectra recorded at $\lambda = 401$~nm at variable delay.}
\end{figure}
Correspondingly, the spectrum (red line) contains a distribution peaked at zero electron kinetic energy, and two sharp peaks at about 15,000 and 16,300~cm$^{-1}$. These electron energies correspond to the photoionization of the Rb atomic 6p-state (red and black dotted lines) and resonance-enhanced two-photon ionization of the Rb-He droplet complex (blue dotted line), respectively. In the case of photoionization with one single laser pulse, only the two-photon peak is measured (blue line). Fig.~\ref{fig:PES} b) shows selected spectra around the two electron peaks as well as the results of fitting the data with a sum of two gaussian functions. To illustrate the time-evolution of the photoelectron energies more clearly, all measured spectra are represented as a contour plot in Fig.~\ref{fig:Contour}. In this representation, the photoelectron feature from two-photon ionization of the Rb-He droplet complex is visible as a bright horizontal band in the upper half of the panel. The Rb atomic 6p photoline appears for $0.3\lesssim\Delta t\lesssim 1.2$~ps as a broadening of the two-photon feature towards lower energies, and develops into a separate band which further shifts downwards for $\Delta t\gtrsim 1.2$~ps. Thus, in the course of the relaxation of the excited Rb-He droplet complex, a fraction of electrons loses about 1175~cm$^{-1}$ of kinetic energy. The time-evolution of the position of this peak is shown in Fig.~\ref{fig:PES} c) with respect to the 6p$_{3/2}$-state of atomic Rb (filled symbols).

The dynamics of electronic relaxation from the excited 6p$\Sigma_{1/2}$-state of the Rb-He droplet complex to the Rb atomic 6p-state is directly correlated with the excited Rb atom detaching from the He droplet surface as seen from the increasing speed of Rb$^+$ ions [Fig.~\ref{fig:transients6p} a)]. The fact that the component of relaxed electrons features a fast energy drop at $0\lesssim \Delta t \lesssim 0.3$~ps which cannot be resolved by the fitting procedure confirms our model assumption of a fast initial rearrangement of the local He environment in response to the excitation of the Rb dopant atom. 


Note that the total loss of potential energy $hc/\lambda-E(\mathrm{6p}_{3/2})=1145$~cm$^{-1}$ occurring during the Rb desorption process exceeds the measured increase of kinetic energy of the Rb atom ($m_\mathrm{Rb}\bar{v}^2/2= 433$~cm$^{-1}$) by about a factor 2.7. Here, $E(\mathrm{6p}_{3/2})=23792.6$~cm$^{-1}$ is the atomic $\mathrm{6p}_{3/2}$-level energy and $\bar{v}=349$~m/s is the asymptotic mean speed of Rb$^+$ ions at $\lambda=401$~nm. Thus, about 2/3 of the released energy is channeled into internal excitations of the droplet by compressing the local surface region of the He droplet and thereby creating a shock wave~\cite{Vangerow:2017,Coppens:2018}. In the present simulation, we assume this to occur promptly corresponding to a sudden shift in the Rb-He droplet distance by 3.3~\AA{}. The rest of the potential energy is then released as kinetic energy of the Rb atom. The experimentally observed drop of electron energy within the first 0.3~ps also indicates fast release of a fraction of the potential energy and a shift in the Rb-He droplet distance, which we estimate to about 1.7~\AA{}. This corresponds to the conversion of 1/3 of the potential energy into internal energy of the droplet. The rest of potential energy is channeled into internal energy of the droplet and into kinetic energy of the Rb atom in the ratio of 1:1. In this slower stage of the desorption dynamics, the released energy is split between the Rb atom and the nanodroplet as in a pseudodiatomic Rb-He$_n$ molecule, where $n\approx 20$. Effectively, the Rb atom ``feels'' the nanodroplet as a group of He atoms in the outer surface layer whose masses are increased by their elastic coupling to He atoms deeper inside the droplet. The best possible convergence of the experiment with the simulation within the PDM would likely be reached by assuming a time-dependent effective mass of the He droplet interacting with the excited Rb atom. However, instead of introducing such an empirical function into our restricted model, deeper insights into the dynamics of this kind of system should be obtainable from dedicated three-dimensional dynamics simulations~\cite{Hernando:2012,Vangerow:2017,Coppens:2018}.


\section{Conclusion}
In conclusion, we have presented a detailed analysis of the shapes of the speed distributions of Rb$^+$ and [RbHe]$^+$ photoions generated by pump-probe photoionization of Rb-doped He nanodroplets. The observed characteristics of the speed distributions are interpreted in terms of coherent wave packet propagation along PECs of the Rb-He droplet complex within the pseudo-diatomic model (PDM). The good agreement of the measured and the simulated evolution of the first to third statistical moments of the distributions confirms that in the dissociation following laser excitation, this  nanometer-sized complex behaves as a quantum object even on the time scale of many picoseconds. Time-resolved photoelectron spectra confirm the direct correspondence of the relaxation of electron binding energy and the motion of escaping Rb atoms. Furthermore, they reveal the branching of energy flow into the motion of the Rb atom and the internal excitations of the He nanodroplet.  These measurements will serve as a reference for future time-resolved studies of similar nanosystems, where the photodynamics are generally more complex and relaxation pathways are harder to disentangle~\cite{Masson:2010,Masson:2014}.

\begin{acknowledgments}
M.M. acknowledges support by Deutsche Forschungsgemeinschaft (project MU 2347/6-1 and IRTG 2079), as well as Aarhus University Research Foundation for a Guest Professorship to A.V.B. N.V.D. and A.V.B. acknowledge support by the Russian Science Foundation (Grant \textnumero 16-13-10024).
\end{acknowledgments}

%


\begin{thebibliography}{56}%
	\makeatletter
	\providecommand \@ifxundefined [1]{%
		\@ifx{#1\undefined}
	}%
	\providecommand \@ifnum [1]{%
		\ifnum #1\expandafter \@firstoftwo
		\else \expandafter \@secondoftwo
		\fi
	}%
	\providecommand \@ifx [1]{%
		\ifx #1\expandafter \@firstoftwo
		\else \expandafter \@secondoftwo
		\fi
	}%
	\providecommand \natexlab [1]{#1}%
	\providecommand \enquote  [1]{``#1''}%
	\providecommand \bibnamefont  [1]{#1}%
	\providecommand \bibfnamefont [1]{#1}%
	\providecommand \citenamefont [1]{#1}%
	\providecommand \href@noop [0]{\@secondoftwo}%
	\providecommand \href [0]{\begingroup \@sanitize@url \@href}%
	\providecommand \@href[1]{\@@startlink{#1}\@@href}%
	\providecommand \@@href[1]{\endgroup#1\@@endlink}%
	\providecommand \@sanitize@url [0]{\catcode `\\12\catcode `\$12\catcode
		`\&12\catcode `\#12\catcode `\^12\catcode `\_12\catcode `\%12\relax}%
	\providecommand \@@startlink[1]{}%
	\providecommand \@@endlink[0]{}%
	\providecommand \url  [0]{\begingroup\@sanitize@url \@url }%
	\providecommand \@url [1]{\endgroup\@href {#1}{\urlprefix }}%
	\providecommand \urlprefix  [0]{URL }%
	\providecommand \Eprint [0]{\href }%
	\providecommand \doibase [0]{http://dx.doi.org/}%
	\providecommand \selectlanguage [0]{\@gobble}%
	\providecommand \bibinfo  [0]{\@secondoftwo}%
	\providecommand \bibfield  [0]{\@secondoftwo}%
	\providecommand \translation [1]{[#1]}%
	\providecommand \BibitemOpen [0]{}%
	\providecommand \bibitemStop [0]{}%
	\providecommand \bibitemNoStop [0]{.\EOS\space}%
	\providecommand \EOS [0]{\spacefactor3000\relax}%
	\providecommand \BibitemShut  [1]{\csname bibitem#1\endcsname}%
	\let\auto@bib@innerbib\@empty
	\bibitem [{\citenamefont {Stienkemeier}\ and\ \citenamefont
		{Lehmann}(2006)}]{Stienkemeier:2006}%
	\BibitemOpen
	\bibfield  {author} {\bibinfo {author} {\bibfnamefont {F.}~\bibnamefont
			{Stienkemeier}}\ and\ \bibinfo {author} {\bibfnamefont {K.}~\bibnamefont
			{Lehmann}},\ }\href@noop {} {\bibfield  {journal} {\bibinfo  {journal}
			{J.~Phys.~B}\ }\textbf {\bibinfo {volume} {39}},\ \bibinfo {pages} {R127}
		(\bibinfo {year} {2006})}\BibitemShut {NoStop}%
	\bibitem [{\citenamefont {Mudrich}\ and\ \citenamefont
		{Stienkemeier}(2014)}]{Mudrich:2014}%
	\BibitemOpen
	\bibfield  {author} {\bibinfo {author} {\bibfnamefont {M.}~\bibnamefont
			{Mudrich}}\ and\ \bibinfo {author} {\bibfnamefont {F.}~\bibnamefont
			{Stienkemeier}},\ }\href@noop {} {\bibfield  {journal} {\bibinfo  {journal}
			{Int. Rev. Phys. Chem.}\ }\textbf {\bibinfo {volume} {33}},\ \bibinfo {pages}
		{301} (\bibinfo {year} {2014})}\BibitemShut {NoStop}%
	\bibitem [{\citenamefont {Ziemkiewicz}\ \emph {et~al.}(2015)\citenamefont
		{Ziemkiewicz}, \citenamefont {Neumark},\ and\ \citenamefont
		{Gessner}}]{Ziemkiewicz:2015}%
	\BibitemOpen
	\bibfield  {author} {\bibinfo {author} {\bibfnamefont {M.~P.}\ \bibnamefont
			{Ziemkiewicz}}, \bibinfo {author} {\bibfnamefont {D.~M.}\ \bibnamefont
			{Neumark}}, \ and\ \bibinfo {author} {\bibfnamefont {O.}~\bibnamefont
			{Gessner}},\ }\href@noop {} {\bibfield  {journal} {\bibinfo  {journal} {Int.
				Rev. Phys. Chem.}\ }\textbf {\bibinfo {volume} {34}},\ \bibinfo {pages} {239}
		(\bibinfo {year} {2015})}\BibitemShut {NoStop}%
	\bibitem [{\citenamefont {Ancilotto}\ \emph {et~al.}(2017)\citenamefont
		{Ancilotto}, \citenamefont {Barranco}, \citenamefont {Coppens}, \citenamefont
		{Eloranta}, \citenamefont {Halberstadt}, \citenamefont {Hernando},
		\citenamefont {Mateo},\ and\ \citenamefont {Pi}}]{Ancilotto:2017}%
	\BibitemOpen
	\bibfield  {author} {\bibinfo {author} {\bibfnamefont {F.}~\bibnamefont
			{Ancilotto}}, \bibinfo {author} {\bibfnamefont {M.}~\bibnamefont {Barranco}},
		\bibinfo {author} {\bibfnamefont {F.}~\bibnamefont {Coppens}}, \bibinfo
		{author} {\bibfnamefont {J.}~\bibnamefont {Eloranta}}, \bibinfo {author}
		{\bibfnamefont {N.}~\bibnamefont {Halberstadt}}, \bibinfo {author}
		{\bibfnamefont {A.}~\bibnamefont {Hernando}}, \bibinfo {author}
		{\bibfnamefont {D.}~\bibnamefont {Mateo}}, \ and\ \bibinfo {author}
		{\bibfnamefont {M.}~\bibnamefont {Pi}},\ }\href@noop {} {\bibfield  {journal}
		{\bibinfo  {journal} {Int. Rev. Phys. Chem.}\ }\textbf {\bibinfo {volume}
			{36}},\ \bibinfo {pages} {621} (\bibinfo {year} {2017})}\BibitemShut
	{NoStop}%
	\bibitem [{\citenamefont {Shepperson}\ \emph {et~al.}(2017)\citenamefont
		{Shepperson}, \citenamefont {S\o{}ndergaard}, \citenamefont {Christiansen},
		\citenamefont {Kaczmarczyk}, \citenamefont {Zillich}, \citenamefont
		{Lemeshko},\ and\ \citenamefont {Stapelfeldt}}]{Shepperson:2017}%
	\BibitemOpen
	\bibfield  {author} {\bibinfo {author} {\bibfnamefont {B.}~\bibnamefont
			{Shepperson}}, \bibinfo {author} {\bibfnamefont {A.~A.}\ \bibnamefont
			{S\o{}ndergaard}}, \bibinfo {author} {\bibfnamefont {L.}~\bibnamefont
			{Christiansen}}, \bibinfo {author} {\bibfnamefont {J.}~\bibnamefont
			{Kaczmarczyk}}, \bibinfo {author} {\bibfnamefont {R.~E.}\ \bibnamefont
			{Zillich}}, \bibinfo {author} {\bibfnamefont {M.}~\bibnamefont {Lemeshko}}, \
		and\ \bibinfo {author} {\bibfnamefont {H.}~\bibnamefont {Stapelfeldt}},\
	}\href@noop {} {\bibfield  {journal} {\bibinfo  {journal} {Phys. Rev. Lett.}\
		}\textbf {\bibinfo {volume} {118}},\ \bibinfo {pages} {203203} (\bibinfo
		{year} {2017})}\BibitemShut {NoStop}%
	\bibitem [{\citenamefont {Loginov}\ and\ \citenamefont
		{Drabbels}(2007)}]{Loginov:2007}%
	\BibitemOpen
	\bibfield  {author} {\bibinfo {author} {\bibfnamefont {E.}~\bibnamefont
			{Loginov}}\ and\ \bibinfo {author} {\bibfnamefont {M.}~\bibnamefont
			{Drabbels}},\ }\href@noop {} {\bibfield  {journal} {\bibinfo  {journal} {J.
				Phys. Chem. A}\ }\textbf {\bibinfo {volume} {111}},\ \bibinfo {pages} {7504}
		(\bibinfo {year} {2007})}\BibitemShut {NoStop}%
	\bibitem [{\citenamefont {Fechner}\ \emph {et~al.}(2012)\citenamefont
		{Fechner}, \citenamefont {Gr{\"u}ner}, \citenamefont {Sieg}, \citenamefont
		{Callegari}, \citenamefont {Ancilotto}, \citenamefont {Stienkemeier},\ and\
		\citenamefont {Mudrich}}]{Fechner:2012}%
	\BibitemOpen
	\bibfield  {author} {\bibinfo {author} {\bibfnamefont {L.}~\bibnamefont
			{Fechner}}, \bibinfo {author} {\bibfnamefont {B.}~\bibnamefont {Gr{\"u}ner}},
		\bibinfo {author} {\bibfnamefont {A.}~\bibnamefont {Sieg}}, \bibinfo {author}
		{\bibfnamefont {C.}~\bibnamefont {Callegari}}, \bibinfo {author}
		{\bibfnamefont {F.}~\bibnamefont {Ancilotto}}, \bibinfo {author}
		{\bibfnamefont {F.}~\bibnamefont {Stienkemeier}}, \ and\ \bibinfo {author}
		{\bibfnamefont {M.}~\bibnamefont {Mudrich}},\ }\href {\doibase
		10.1039/C2CP22749E} {\bibfield  {journal} {\bibinfo  {journal} {Phys. Chem.
				Chem. Phys.}\ }\textbf {\bibinfo {volume} {14}},\ \bibinfo {pages} {3843}
		(\bibinfo {year} {2012})}\BibitemShut {NoStop}%
	\bibitem [{\citenamefont {Kautsch}\ \emph {et~al.}(2013)\citenamefont
		{Kautsch}, \citenamefont {Koch},\ and\ \citenamefont {Ernst}}]{Kautsch:2013}%
	\BibitemOpen
	\bibfield  {author} {\bibinfo {author} {\bibfnamefont {A.}~\bibnamefont
			{Kautsch}}, \bibinfo {author} {\bibfnamefont {M.}~\bibnamefont {Koch}}, \
		and\ \bibinfo {author} {\bibfnamefont {W.~E.}\ \bibnamefont {Ernst}},\
	}\href@noop {} {\bibfield  {journal} {\bibinfo  {journal} {J. Phys. Chem. A}\
		}\textbf {\bibinfo {volume} {117}},\ \bibinfo {pages} {9621} (\bibinfo {year}
		{2013})}\BibitemShut {NoStop}%
	\bibitem [{\citenamefont {Lindebner}\ \emph {et~al.}(2014)\citenamefont
		{Lindebner}, \citenamefont {Kautsch}, \citenamefont {Koch},\ and\
		\citenamefont {Ernst}}]{Lindebner:2014}%
	\BibitemOpen
	\bibfield  {author} {\bibinfo {author} {\bibfnamefont {F.}~\bibnamefont
			{Lindebner}}, \bibinfo {author} {\bibfnamefont {A.}~\bibnamefont {Kautsch}},
		\bibinfo {author} {\bibfnamefont {M.}~\bibnamefont {Koch}}, \ and\ \bibinfo
		{author} {\bibfnamefont {W.~E.}\ \bibnamefont {Ernst}},\ }\href {\doibase
		http://dx.doi.org/10.1016/j.ijms.2013.12.022} {\bibfield  {journal} {\bibinfo
			{journal} {Int. J. Mass Spectrom.}\ }\textbf {\bibinfo {volume} {365-366}},\
		\bibinfo {pages} {255} (\bibinfo {year} {2014})}\BibitemShut {NoStop}%
	\bibitem [{\citenamefont {Braun}\ and\ \citenamefont
		{Drabbels}(2004)}]{Braun:2004}%
	\BibitemOpen
	\bibfield  {author} {\bibinfo {author} {\bibfnamefont {A.}~\bibnamefont
			{Braun}}\ and\ \bibinfo {author} {\bibfnamefont {M.}~\bibnamefont
			{Drabbels}},\ }\href {\doibase 10.1103/PhysRevLett.93.253401} {\bibfield
		{journal} {\bibinfo  {journal} {Phys. Rev. Lett.}\ }\textbf {\bibinfo
			{volume} {93}},\ \bibinfo {pages} {253401} (\bibinfo {year}
		{2004})}\BibitemShut {NoStop}%
	\bibitem [{\citenamefont {Kautsch}\ \emph {et~al.}(2015)\citenamefont
		{Kautsch}, \citenamefont {Koch},\ and\ \citenamefont {Ernst}}]{Kautsch:2015}%
	\BibitemOpen
	\bibfield  {author} {\bibinfo {author} {\bibfnamefont {A.}~\bibnamefont
			{Kautsch}}, \bibinfo {author} {\bibfnamefont {M.}~\bibnamefont {Koch}}, \
		and\ \bibinfo {author} {\bibfnamefont {W.~E.}\ \bibnamefont {Ernst}},\ }\href
	{\doibase 10.1039/C5CP01009H} {\bibfield  {journal} {\bibinfo  {journal}
			{Phys. Chem. Chem. Phys.}\ }\textbf {\bibinfo {volume} {17}},\ \bibinfo
		{pages} {12310} (\bibinfo {year} {2015})}\BibitemShut {NoStop}%
	\bibitem [{\citenamefont {Reho}\ \emph
		{et~al.}(2000{\natexlab{a}})\citenamefont {Reho}, \citenamefont {Higgins},
		\citenamefont {Callegari}, \citenamefont {Lehmann},\ and\ \citenamefont
		{Scoles}}]{Reho:2000}%
	\BibitemOpen
	\bibfield  {author} {\bibinfo {author} {\bibfnamefont {J.}~\bibnamefont
			{Reho}}, \bibinfo {author} {\bibfnamefont {J.}~\bibnamefont {Higgins}},
		\bibinfo {author} {\bibfnamefont {C.}~\bibnamefont {Callegari}}, \bibinfo
		{author} {\bibfnamefont {K.~K.}\ \bibnamefont {Lehmann}}, \ and\ \bibinfo
		{author} {\bibfnamefont {G.}~\bibnamefont {Scoles}},\ }\href@noop {}
	{\bibfield  {journal} {\bibinfo  {journal} {J. Chem. Phys.}\ }\textbf
		{\bibinfo {volume} {113}},\ \bibinfo {pages} {9686} (\bibinfo {year}
		{2000}{\natexlab{a}})}\BibitemShut {NoStop}%
	\bibitem [{\citenamefont {Br{\"u}hl}\ \emph {et~al.}(2001)\citenamefont
		{Br{\"u}hl}, \citenamefont {Trasca},\ and\ \citenamefont
		{Ernst}}]{Bruehl:2001}%
	\BibitemOpen
	\bibfield  {author} {\bibinfo {author} {\bibfnamefont {F.~R.}\ \bibnamefont
			{Br{\"u}hl}}, \bibinfo {author} {\bibfnamefont {R.~A.}\ \bibnamefont
			{Trasca}}, \ and\ \bibinfo {author} {\bibfnamefont {W.~E.}\ \bibnamefont
			{Ernst}},\ }\href {\doibase 10.1063/1.1410118} {\bibfield  {journal}
		{\bibinfo  {journal} {J. Chem. Phys.}\ }\textbf {\bibinfo {volume} {115}},\
		\bibinfo {pages} {10220} (\bibinfo {year} {2001})}\BibitemShut {NoStop}%
	\bibitem [{\citenamefont {Droppelmann}\ \emph {et~al.}(2004)\citenamefont
		{Droppelmann}, \citenamefont {B{\"u}nermann}, \citenamefont {Schulz},\ and\
		\citenamefont {Stienkemeier}}]{Droppelmann:2004}%
	\BibitemOpen
	\bibfield  {author} {\bibinfo {author} {\bibfnamefont {G.}~\bibnamefont
			{Droppelmann}}, \bibinfo {author} {\bibfnamefont {O.}~\bibnamefont
			{B{\"u}nermann}}, \bibinfo {author} {\bibfnamefont {C.~P.}\ \bibnamefont
			{Schulz}}, \ and\ \bibinfo {author} {\bibfnamefont {F.}~\bibnamefont
			{Stienkemeier}},\ }\href@noop {} {\bibfield  {journal} {\bibinfo  {journal}
			{Phys. Rev. Lett.}\ }\textbf {\bibinfo {volume} {93}},\ \bibinfo {pages}
		{023402} (\bibinfo {year} {2004})}\BibitemShut {NoStop}%
	\bibitem [{\citenamefont {Leino}\ \emph {et~al.}(2011)\citenamefont {Leino},
		\citenamefont {Viel},\ and\ \citenamefont {Zillich}}]{Leino:2011}%
	\BibitemOpen
	\bibfield  {author} {\bibinfo {author} {\bibfnamefont {M.}~\bibnamefont
			{Leino}}, \bibinfo {author} {\bibfnamefont {A.}~\bibnamefont {Viel}}, \ and\
		\bibinfo {author} {\bibfnamefont {R.~E.}\ \bibnamefont {Zillich}},\ }\href
	{\doibase 10.1063/1.3528936} {\bibfield  {journal} {\bibinfo  {journal} {J.
				Chem. Phys.}\ }\textbf {\bibinfo {volume} {134}},\ \bibinfo {eid} {024316}
		(\bibinfo {year} {2011})}\BibitemShut {NoStop}%
	\bibitem [{\citenamefont {G{\"o}de}\ \emph {et~al.}(2013)\citenamefont
		{G{\"o}de}, \citenamefont {Irsig}, \citenamefont {Tiggesb{\"a}umker},\ and\
		\citenamefont {Meiwes-Broer}}]{Goede:2013}%
	\BibitemOpen
	\bibfield  {author} {\bibinfo {author} {\bibfnamefont {S.}~\bibnamefont
			{G{\"o}de}}, \bibinfo {author} {\bibfnamefont {R.}~\bibnamefont {Irsig}},
		\bibinfo {author} {\bibfnamefont {J.}~\bibnamefont {Tiggesb{\"a}umker}}, \
		and\ \bibinfo {author} {\bibfnamefont {K.-H.}\ \bibnamefont {Meiwes-Broer}},\
	}\href {http://stacks.iop.org/1367-2630/15/i=1/a=015026} {\bibfield
		{journal} {\bibinfo  {journal} {New J. Phys.}\ }\textbf {\bibinfo {volume}
			{15}},\ \bibinfo {pages} {015026} (\bibinfo {year} {2013})}\BibitemShut
	{NoStop}%
	\bibitem [{\citenamefont {Lackner}\ and\ \citenamefont
		{Ernst}(2018)}]{Lackner:2018}%
	\BibitemOpen
	\bibfield  {author} {\bibinfo {author} {\bibfnamefont {F.}~\bibnamefont
			{Lackner}}\ and\ \bibinfo {author} {\bibfnamefont {W.~E.}\ \bibnamefont
			{Ernst}},\ }\href@noop {} {\bibfield  {journal} {\bibinfo  {journal} {J.
				Phys. Chem. Lett.}\ }\textbf {\bibinfo {volume} {9}},\ \bibinfo {pages}
		{3561} (\bibinfo {year} {2018})}\BibitemShut {NoStop}%
	\bibitem [{\citenamefont {Brauer}\ \emph {et~al.}(2013)\citenamefont {Brauer},
		\citenamefont {Smolarek}, \citenamefont {Loginov}, \citenamefont {Mateo},
		\citenamefont {Hernando}, \citenamefont {Pi}, \citenamefont {Barranco},
		\citenamefont {Buma},\ and\ \citenamefont {Drabbels}}]{Brauer:2013}%
	\BibitemOpen
	\bibfield  {author} {\bibinfo {author} {\bibfnamefont {N.~B.}\ \bibnamefont
			{Brauer}}, \bibinfo {author} {\bibfnamefont {S.}~\bibnamefont {Smolarek}},
		\bibinfo {author} {\bibfnamefont {E.}~\bibnamefont {Loginov}}, \bibinfo
		{author} {\bibfnamefont {D.}~\bibnamefont {Mateo}}, \bibinfo {author}
		{\bibfnamefont {A.}~\bibnamefont {Hernando}}, \bibinfo {author}
		{\bibfnamefont {M.}~\bibnamefont {Pi}}, \bibinfo {author} {\bibfnamefont
			{M.}~\bibnamefont {Barranco}}, \bibinfo {author} {\bibfnamefont {W.~J.}\
			\bibnamefont {Buma}}, \ and\ \bibinfo {author} {\bibfnamefont
			{M.}~\bibnamefont {Drabbels}},\ }\href {\doibase
		10.1103/PhysRevLett.111.153002} {\bibfield  {journal} {\bibinfo  {journal}
			{Phys. Rev. Lett.}\ }\textbf {\bibinfo {volume} {111}},\ \bibinfo {pages}
		{153002} (\bibinfo {year} {2013})}\BibitemShut {NoStop}%
	\bibitem [{\citenamefont {Koch}\ \emph {et~al.}(2014)\citenamefont {Koch},
		\citenamefont {Kautsch}, \citenamefont {Lackner},\ and\ \citenamefont
		{Ernst}}]{Koch:2014}%
	\BibitemOpen
	\bibfield  {author} {\bibinfo {author} {\bibfnamefont {M.}~\bibnamefont
			{Koch}}, \bibinfo {author} {\bibfnamefont {A.}~\bibnamefont {Kautsch}},
		\bibinfo {author} {\bibfnamefont {F.}~\bibnamefont {Lackner}}, \ and\
		\bibinfo {author} {\bibfnamefont {W.~E.}\ \bibnamefont {Ernst}},\ }\href@noop
	{} {\bibfield  {journal} {\bibinfo  {journal} {J. Phys. Chem. A}\ }\textbf
		{\bibinfo {volume} {118}},\ \bibinfo {pages} {8373} (\bibinfo {year}
		{2014})}\BibitemShut {NoStop}%
	\bibitem [{\citenamefont {Seki}\ \emph {et~al.}(2017)\citenamefont {Seki},
		\citenamefont {Takayanagi},\ and\ \citenamefont {Shiga}}]{Takayanagi:2017}%
	\BibitemOpen
	\bibfield  {author} {\bibinfo {author} {\bibfnamefont {Y.}~\bibnamefont
			{Seki}}, \bibinfo {author} {\bibfnamefont {T.}~\bibnamefont {Takayanagi}}, \
		and\ \bibinfo {author} {\bibfnamefont {M.}~\bibnamefont {Shiga}},\ }\href
	{\doibase 10.1039/C7CP00888K} {\bibfield  {journal} {\bibinfo  {journal}
			{Phys. Chem. Chem. Phys.}\ }\textbf {\bibinfo {volume} {19}},\ \bibinfo
		{pages} {13798} (\bibinfo {year} {2017})}\BibitemShut {NoStop}%
	\bibitem [{\citenamefont {Stienkemeier}\ \emph {et~al.}(1996)\citenamefont
		{Stienkemeier}, \citenamefont {Higgins}, \citenamefont {Callegari},
		\citenamefont {Kanorsky}, \citenamefont {Ernst},\ and\ \citenamefont
		{Scoles}}]{Stienkemeier:1996}%
	\BibitemOpen
	\bibfield  {author} {\bibinfo {author} {\bibfnamefont {F.}~\bibnamefont
			{Stienkemeier}}, \bibinfo {author} {\bibfnamefont {J.}~\bibnamefont
			{Higgins}}, \bibinfo {author} {\bibfnamefont {C.}~\bibnamefont {Callegari}},
		\bibinfo {author} {\bibfnamefont {S.~I.}\ \bibnamefont {Kanorsky}}, \bibinfo
		{author} {\bibfnamefont {W.~E.}\ \bibnamefont {Ernst}}, \ and\ \bibinfo
		{author} {\bibfnamefont {G.}~\bibnamefont {Scoles}},\ }\href@noop {}
	{\bibfield  {journal} {\bibinfo  {journal} {Z. Phys. D}\ }\textbf {\bibinfo
			{volume} {38}},\ \bibinfo {pages} {253} (\bibinfo {year} {1996})}\BibitemShut
	{NoStop}%
	\bibitem [{\citenamefont {Barranco}\ \emph {et~al.}(2006)\citenamefont
		{Barranco}, \citenamefont {Guardiola}, \citenamefont {Hern{\'a}ndez},
		\citenamefont {Mayol}, \citenamefont {Navarro},\ and\ \citenamefont
		{Pi}}]{Barranco:2006}%
	\BibitemOpen
	\bibfield  {author} {\bibinfo {author} {\bibfnamefont {M.}~\bibnamefont
			{Barranco}}, \bibinfo {author} {\bibfnamefont {R.}~\bibnamefont {Guardiola}},
		\bibinfo {author} {\bibfnamefont {S.}~\bibnamefont {Hern{\'a}ndez}}, \bibinfo
		{author} {\bibfnamefont {R.}~\bibnamefont {Mayol}}, \bibinfo {author}
		{\bibfnamefont {J.}~\bibnamefont {Navarro}}, \ and\ \bibinfo {author}
		{\bibfnamefont {M.}~\bibnamefont {Pi}},\ }\href@noop {} {\bibfield  {journal}
		{\bibinfo  {journal} {J. Low Temp. Phys.}\ }\textbf {\bibinfo {volume}
			{142}},\ \bibinfo {pages} {1} (\bibinfo {year} {2006})}\BibitemShut {NoStop}%
	\bibitem [{\citenamefont {Aub{\"o}ck}\ \emph {et~al.}(2008)\citenamefont
		{Aub{\"o}ck}, \citenamefont {Nagl}, \citenamefont {Callegari},\ and\
		\citenamefont {Ernst}}]{Auboeck:2008}%
	\BibitemOpen
	\bibfield  {author} {\bibinfo {author} {\bibfnamefont {G.}~\bibnamefont
			{Aub{\"o}ck}}, \bibinfo {author} {\bibfnamefont {J.}~\bibnamefont {Nagl}},
		\bibinfo {author} {\bibfnamefont {C.}~\bibnamefont {Callegari}}, \ and\
		\bibinfo {author} {\bibfnamefont {W.~E.}\ \bibnamefont {Ernst}},\ }\href@noop
	{} {\bibfield  {journal} {\bibinfo  {journal} {Phys. Rev. Lett.}\ }\textbf
		{\bibinfo {volume} {101}},\ \bibinfo {pages} {035301} (\bibinfo {year}
		{2008})}\BibitemShut {NoStop}%
	\bibitem [{\citenamefont {Theisen}\ \emph {et~al.}(2011)\citenamefont
		{Theisen}, \citenamefont {Lackner},\ and\ \citenamefont
		{Ernst}}]{Theisen:2011}%
	\BibitemOpen
	\bibfield  {author} {\bibinfo {author} {\bibfnamefont {M.}~\bibnamefont
			{Theisen}}, \bibinfo {author} {\bibfnamefont {F.}~\bibnamefont {Lackner}}, \
		and\ \bibinfo {author} {\bibfnamefont {W.~E.}\ \bibnamefont {Ernst}},\
	}\href@noop {} {\bibfield  {journal} {\bibinfo  {journal} {J. Phys. Chem. A}\
		}\textbf {\bibinfo {volume} {115}},\ \bibinfo {pages} {7005} (\bibinfo {year}
		{2011})}\BibitemShut {NoStop}%
	\bibitem [{\citenamefont {Takayanagi}\ and\ \citenamefont
		{Shiga}(2004)}]{Takayanagi:2004}%
	\BibitemOpen
	\bibfield  {author} {\bibinfo {author} {\bibfnamefont {T.}~\bibnamefont
			{Takayanagi}}\ and\ \bibinfo {author} {\bibfnamefont {M.}~\bibnamefont
			{Shiga}},\ }\href@noop {} {\bibfield  {journal} {\bibinfo  {journal} {Phys.
				Chem. Chem. Phys.}\ }\textbf {\bibinfo {volume} {6}},\ \bibinfo {pages}
		{3241} (\bibinfo {year} {2004})}\BibitemShut {NoStop}%
	\bibitem [{\citenamefont {Hernando}\ \emph {et~al.}(2012)\citenamefont
		{Hernando}, \citenamefont {Barranco}, \citenamefont {Pi}, \citenamefont
		{Loginov}, \citenamefont {Langlet},\ and\ \citenamefont
		{Drabbels}}]{Hernando:2012}%
	\BibitemOpen
	\bibfield  {author} {\bibinfo {author} {\bibfnamefont {A.}~\bibnamefont
			{Hernando}}, \bibinfo {author} {\bibfnamefont {M.}~\bibnamefont {Barranco}},
		\bibinfo {author} {\bibfnamefont {M.}~\bibnamefont {Pi}}, \bibinfo {author}
		{\bibfnamefont {E.}~\bibnamefont {Loginov}}, \bibinfo {author} {\bibfnamefont
			{M.}~\bibnamefont {Langlet}}, \ and\ \bibinfo {author} {\bibfnamefont
			{M.}~\bibnamefont {Drabbels}},\ }\href {\doibase 10.1039/C2CP23526A}
	{\bibfield  {journal} {\bibinfo  {journal} {Phys. Chem. Chem. Phys.}\
		}\textbf {\bibinfo {volume} {14}},\ \bibinfo {pages} {3996} (\bibinfo {year}
		{2012})}\BibitemShut {NoStop}%
	\bibitem [{\citenamefont {von Vangerow}\ \emph {et~al.}(2014)\citenamefont {von
			Vangerow}, \citenamefont {Sieg}, \citenamefont {Stienkemeier}, \citenamefont
		{Mudrich}, \citenamefont {Leal}, \citenamefont {Mateo}, \citenamefont
		{Hernando}, \citenamefont {Barranco},\ and\ \citenamefont
		{Pi}}]{Vangerow:2014}%
	\BibitemOpen
	\bibfield  {author} {\bibinfo {author} {\bibfnamefont {J.}~\bibnamefont {von
				Vangerow}}, \bibinfo {author} {\bibfnamefont {A.}~\bibnamefont {Sieg}},
		\bibinfo {author} {\bibfnamefont {F.}~\bibnamefont {Stienkemeier}}, \bibinfo
		{author} {\bibfnamefont {M.}~\bibnamefont {Mudrich}}, \bibinfo {author}
		{\bibfnamefont {A.}~\bibnamefont {Leal}}, \bibinfo {author} {\bibfnamefont
			{D.}~\bibnamefont {Mateo}}, \bibinfo {author} {\bibfnamefont
			{A.}~\bibnamefont {Hernando}}, \bibinfo {author} {\bibfnamefont
			{M.}~\bibnamefont {Barranco}}, \ and\ \bibinfo {author} {\bibfnamefont
			{M.}~\bibnamefont {Pi}},\ }\href@noop {} {\bibfield  {journal} {\bibinfo
			{journal} {J. Phys. Chem. A}\ }\textbf {\bibinfo {volume} {118}},\ \bibinfo
		{pages} {6604} (\bibinfo {year} {2014})}\BibitemShut {NoStop}%
	\bibitem [{\citenamefont {von Vangerow}\ \emph {et~al.}(2015)\citenamefont {von
			Vangerow}, \citenamefont {John}, \citenamefont {Stienkemeier},\ and\
		\citenamefont {Mudrich}}]{Vangerow:2015}%
	\BibitemOpen
	\bibfield  {author} {\bibinfo {author} {\bibfnamefont {J.}~\bibnamefont {von
				Vangerow}}, \bibinfo {author} {\bibfnamefont {O.}~\bibnamefont {John}},
		\bibinfo {author} {\bibfnamefont {F.}~\bibnamefont {Stienkemeier}}, \ and\
		\bibinfo {author} {\bibfnamefont {M.}~\bibnamefont {Mudrich}},\ }\href@noop
	{} {\bibfield  {journal} {\bibinfo  {journal} {J. Chem. Phys.}\ }\textbf
		{\bibinfo {volume} {143}},\ \bibinfo {pages} {034302} (\bibinfo {year}
		{2015})}\BibitemShut {NoStop}%
	\bibitem [{\citenamefont {von Vangerow}\ \emph {et~al.}(2017)\citenamefont {von
			Vangerow}, \citenamefont {Coppens}, \citenamefont {Leal}, \citenamefont {Pi},
		\citenamefont {Barranco}, \citenamefont {Halberstadt}, \citenamefont
		{Stienkemeier},\ and\ \citenamefont {Mudrich}}]{Vangerow:2017}%
	\BibitemOpen
	\bibfield  {author} {\bibinfo {author} {\bibfnamefont {J.}~\bibnamefont {von
				Vangerow}}, \bibinfo {author} {\bibfnamefont {F.}~\bibnamefont {Coppens}},
		\bibinfo {author} {\bibfnamefont {A.}~\bibnamefont {Leal}}, \bibinfo {author}
		{\bibfnamefont {M.}~\bibnamefont {Pi}}, \bibinfo {author} {\bibfnamefont
			{M.}~\bibnamefont {Barranco}}, \bibinfo {author} {\bibfnamefont
			{N.}~\bibnamefont {Halberstadt}}, \bibinfo {author} {\bibfnamefont
			{F.}~\bibnamefont {Stienkemeier}}, \ and\ \bibinfo {author} {\bibfnamefont
			{M.}~\bibnamefont {Mudrich}},\ }\href@noop {} {\bibfield  {journal} {\bibinfo
			{journal} {J. Phys. Chem. Lett.}\ }\textbf {\bibinfo {volume} {8}},\
		\bibinfo {pages} {307} (\bibinfo {year} {2017})}\BibitemShut {NoStop}%
	\bibitem [{\citenamefont {Coppens}\ \emph {et~al.}(2018)\citenamefont
		{Coppens}, \citenamefont {von Vangerow}, \citenamefont {Barranco},
		\citenamefont {Halberstadt}, \citenamefont {Stienkemeier}, \citenamefont
		{Pi},\ and\ \citenamefont {Mudrich}}]{Coppens:2018}%
	\BibitemOpen
	\bibfield  {author} {\bibinfo {author} {\bibfnamefont {F.}~\bibnamefont
			{Coppens}}, \bibinfo {author} {\bibfnamefont {J.}~\bibnamefont {von
				Vangerow}}, \bibinfo {author} {\bibfnamefont {M.}~\bibnamefont {Barranco}},
		\bibinfo {author} {\bibfnamefont {N.}~\bibnamefont {Halberstadt}}, \bibinfo
		{author} {\bibfnamefont {F.}~\bibnamefont {Stienkemeier}}, \bibinfo {author}
		{\bibfnamefont {M.}~\bibnamefont {Pi}}, \ and\ \bibinfo {author}
		{\bibfnamefont {M.}~\bibnamefont {Mudrich}},\ }\href@noop {} {\bibfield
		{journal} {\bibinfo  {journal} {Phys. Chem. Chem. Phys.}\ }\textbf {\bibinfo
			{volume} {20}},\ \bibinfo {pages} {9309} (\bibinfo {year}
		{2018})}\BibitemShut {NoStop}%
	\bibitem [{\citenamefont {Loginov}\ and\ \citenamefont
		{Drabbels}(2011)}]{LoginovPRL:2011}%
	\BibitemOpen
	\bibfield  {author} {\bibinfo {author} {\bibfnamefont {E.}~\bibnamefont
			{Loginov}}\ and\ \bibinfo {author} {\bibfnamefont {M.}~\bibnamefont
			{Drabbels}},\ }\href {\doibase 10.1103/PhysRevLett.106.083401} {\bibfield
		{journal} {\bibinfo  {journal} {Phys. Rev. Lett.}\ }\textbf {\bibinfo
			{volume} {106}},\ \bibinfo {pages} {083401} (\bibinfo {year}
		{2011})}\BibitemShut {NoStop}%
	\bibitem [{\citenamefont {Callegari}\ and\ \citenamefont
		{Ancilotto}(2011)}]{Callegari:2011}%
	\BibitemOpen
	\bibfield  {author} {\bibinfo {author} {\bibfnamefont {C.}~\bibnamefont
			{Callegari}}\ and\ \bibinfo {author} {\bibfnamefont {F.}~\bibnamefont
			{Ancilotto}},\ }\href@noop {} {\bibfield  {journal} {\bibinfo  {journal} {J.
				Phys. Chem. A}\ }\textbf {\bibinfo {volume} {115}},\ \bibinfo {pages} {6789}
		(\bibinfo {year} {2011})}\BibitemShut {NoStop}%
	\bibitem [{\citenamefont {Eppink}\ and\ \citenamefont
		{Parker}(1997)}]{Eppink:1997}%
	\BibitemOpen
	\bibfield  {author} {\bibinfo {author} {\bibfnamefont {A.~T. J.~B.}\
			\bibnamefont {Eppink}}\ and\ \bibinfo {author} {\bibfnamefont {D.~H.}\
			\bibnamefont {Parker}},\ }\href@noop {} {\bibfield  {journal} {\bibinfo
			{journal} {Rev. Sci. Instrum.}\ }\textbf {\bibinfo {volume} {68}},\ \bibinfo
		{pages} {3477} (\bibinfo {year} {1997})}\BibitemShut {NoStop}%
	\bibitem [{\citenamefont {Dick}(2014)}]{Dick:2014}%
	\BibitemOpen
	\bibfield  {author} {\bibinfo {author} {\bibfnamefont {B.}~\bibnamefont
			{Dick}},\ }\href@noop {} {\bibfield  {journal} {\bibinfo  {journal} {Phys.
				Chem. Chem. Phys.}\ }\textbf {\bibinfo {volume} {16}},\ \bibinfo {pages}
		{570} (\bibinfo {year} {2014})}\BibitemShut {NoStop}%
	\bibitem [{\citenamefont {Mudholkar}\ and\ \citenamefont
		{Hutson}(2000{\natexlab{a}})}]{mudholkar2000epsilon}%
	\BibitemOpen
	\bibfield  {author} {\bibinfo {author} {\bibfnamefont {G.~S.}\ \bibnamefont
			{Mudholkar}}\ and\ \bibinfo {author} {\bibfnamefont {A.~D.}\ \bibnamefont
			{Hutson}},\ }\href@noop {} {\bibfield  {journal} {\bibinfo  {journal} {J.
				Stat. Plan. Inference}\ }\textbf {\bibinfo {volume} {83}},\ \bibinfo {pages}
		{291} (\bibinfo {year} {2000}{\natexlab{a}})}\BibitemShut {NoStop}%
	\bibitem [{\citenamefont {Feit}\ \emph {et~al.}(1982)\citenamefont {Feit},
		\citenamefont {Fleck},\ and\ \citenamefont {Steiger}}]{Feit:1982}%
	\BibitemOpen
	\bibfield  {author} {\bibinfo {author} {\bibfnamefont {M.}~\bibnamefont
			{Feit}}, \bibinfo {author} {\bibfnamefont {F.}~\bibnamefont {Fleck}}, \ and\
		\bibinfo {author} {\bibfnamefont {A.}~\bibnamefont {Steiger}},\ }\href@noop
	{} {\bibfield  {journal} {\bibinfo  {journal} {J. Comp. Phys.}\ }\textbf
		{\bibinfo {volume} {47}},\ \bibinfo {pages} {412} (\bibinfo {year}
		{1982})}\BibitemShut {NoStop}%
	\bibitem [{\citenamefont {Schmidt}\ and\ \citenamefont
		{Lorenz}(2017)}]{Schmidt:2017}%
	\BibitemOpen
	\bibfield  {author} {\bibinfo {author} {\bibfnamefont {B.}~\bibnamefont
			{Schmidt}}\ and\ \bibinfo {author} {\bibfnamefont {U.}~\bibnamefont
			{Lorenz}},\ }\href@noop {} {\bibfield  {journal} {\bibinfo  {journal}
			{Comput. Phys. Commun.}\ }\textbf {\bibinfo {volume} {213}},\ \bibinfo
		{pages} {223 } (\bibinfo {year} {2017})}\BibitemShut {NoStop}%
	\bibitem [{\citenamefont {Tannor}(2007)}]{TannorBook}%
	\BibitemOpen
	\bibfield  {author} {\bibinfo {author} {\bibfnamefont {D.~J.}\ \bibnamefont
			{Tannor}},\ }\href@noop {} {\emph {\bibinfo {title} {Introduction to Quantum
				Mechanics -- A Time-Dependent Perspective}}}\ (\bibinfo  {publisher}
	{University Science Books},\ \bibinfo {year} {2007})\BibitemShut {NoStop}%
	\bibitem [{\citenamefont {Reho}\ \emph
		{et~al.}(2000{\natexlab{b}})\citenamefont {Reho}, \citenamefont {Higgins},
		\citenamefont {Callegari}, \citenamefont {K.Lehmann},\ and\ \citenamefont
		{Scoles}}]{Reho2:2000}%
	\BibitemOpen
	\bibfield  {author} {\bibinfo {author} {\bibfnamefont {J.}~\bibnamefont
			{Reho}}, \bibinfo {author} {\bibfnamefont {J.}~\bibnamefont {Higgins}},
		\bibinfo {author} {\bibfnamefont {C.}~\bibnamefont {Callegari}}, \bibinfo
		{author} {\bibfnamefont {K.}~\bibnamefont {K.Lehmann}}, \ and\ \bibinfo
		{author} {\bibfnamefont {G.}~\bibnamefont {Scoles}},\ }\href@noop {}
	{\bibfield  {journal} {\bibinfo  {journal} {J. Chem. Phys.}\ }\textbf
		{\bibinfo {volume} {113}},\ \bibinfo {pages} {9694} (\bibinfo {year}
		{2000}{\natexlab{b}})}\BibitemShut {NoStop}%
	\bibitem [{\citenamefont {Koutselos}\ \emph {et~al.}(1990)\citenamefont
		{Koutselos}, \citenamefont {Mason},\ and\ \citenamefont
		{Viehland}}]{Koutselos:1990}%
	\BibitemOpen
	\bibfield  {author} {\bibinfo {author} {\bibfnamefont {A.~D.}\ \bibnamefont
			{Koutselos}}, \bibinfo {author} {\bibfnamefont {E.~A.}\ \bibnamefont
			{Mason}}, \ and\ \bibinfo {author} {\bibfnamefont {L.~A.}\ \bibnamefont
			{Viehland}},\ }\href@noop {} {\bibfield  {journal} {\bibinfo  {journal} {J.
				Chem. Phys.}\ }\textbf {\bibinfo {volume} {93}},\ \bibinfo {pages} {7125}
		(\bibinfo {year} {1990})}\BibitemShut {NoStop}%
	\bibitem [{\citenamefont {Pi}(2014)}]{Pi}%
	\BibitemOpen
	\bibfield  {author} {\bibinfo {author} {\bibfnamefont {M.}~\bibnamefont
			{Pi}},\ }\href@noop {} {\enquote {\bibinfo {title} {Private communication},}\
	} (\bibinfo {year} {2014})\BibitemShut {NoStop}%
	\bibitem [{\citenamefont {LeRoy}(1995)}]{level}%
	\BibitemOpen
	\bibfield  {author} {\bibinfo {author} {\bibfnamefont {R.}~\bibnamefont
			{LeRoy}},\ }\href@noop {} {\enquote {\bibinfo {title} {Chemical physics
				research report},}\ }\bibinfo {howpublished} {University of Waterloo, CP-555,
		1995} (\bibinfo {year} {1995})\BibitemShut {NoStop}%
	\bibitem [{Mat()}]{Mathematica}%
	\BibitemOpen
	\href@noop {} {}\bibinfo {howpublished}
	{\url{https://www.wolfram.com/mathematica/}}\BibitemShut {NoStop}%
	\bibitem [{\citenamefont {Zare}(1972)}]{Zare:1972}%
	\BibitemOpen
	\bibfield  {author} {\bibinfo {author} {\bibfnamefont {R.~N.}\ \bibnamefont
			{Zare}},\ }\href@noop {} {\bibfield  {journal} {\bibinfo  {journal} {Mol.
				Photochem.}\ }\textbf {\bibinfo {volume} {44}},\ \bibinfo {pages} {1}
		(\bibinfo {year} {1972})}\BibitemShut {NoStop}%
	\bibitem [{\citenamefont {Mudholkar}\ and\ \citenamefont
		{Hutson}(2000{\natexlab{b}})}]{Mudholkar:2000}%
	\BibitemOpen
	\bibfield  {author} {\bibinfo {author} {\bibfnamefont {G.~S.}\ \bibnamefont
			{Mudholkar}}\ and\ \bibinfo {author} {\bibfnamefont {A.~D.}\ \bibnamefont
			{Hutson}},\ }\href@noop {} {\bibfield  {journal} {\bibinfo  {journal} {J.
				Statist. Plann. Inference}\ }\textbf {\bibinfo {volume} {83}},\ \bibinfo
		{pages} {291−309} (\bibinfo {year} {2000}{\natexlab{b}})}\BibitemShut
	{NoStop}%
	\bibitem [{\citenamefont {Tannor}\ and\ \citenamefont
		{Rice}(1985)}]{Tannor:1985}%
	\BibitemOpen
	\bibfield  {author} {\bibinfo {author} {\bibfnamefont {D.~J.}\ \bibnamefont
			{Tannor}}\ and\ \bibinfo {author} {\bibfnamefont {S.~A.}\ \bibnamefont
			{Rice}},\ }\href@noop {} {\bibfield  {journal} {\bibinfo  {journal} {J. Chem.
				Phys.}\ }\textbf {\bibinfo {volume} {83}},\ \bibinfo {pages} {5013} (\bibinfo
		{year} {1985})}\BibitemShut {NoStop}%
	\bibitem [{\citenamefont {Zewail}(1994)}]{Zewail:1994}%
	\BibitemOpen
	\bibfield  {author} {\bibinfo {author} {\bibfnamefont {A.}~\bibnamefont
			{Zewail}},\ }\href@noop {} {\emph {\bibinfo {title} {Femtochemistry}}}\
	(\bibinfo  {publisher} {World Scientific, Singapore, 1994},\ \bibinfo {year}
	{1994})\BibitemShut {NoStop}%
	\bibitem [{\citenamefont {Schmidt}\ \emph {et~al.}(2015)\citenamefont
		{Schmidt}, \citenamefont {von Vangerow}, \citenamefont {Stienkemeier},
		\citenamefont {Bogomolov}, \citenamefont {Baklanov}, \citenamefont {Reich},
		\citenamefont {Skomorowski}, \citenamefont {Koch},\ and\ \citenamefont
		{Mudrich}}]{Schmidt:2014}%
	\BibitemOpen
	\bibfield  {author} {\bibinfo {author} {\bibfnamefont {H.}~\bibnamefont
			{Schmidt}}, \bibinfo {author} {\bibfnamefont {J.}~\bibnamefont {von
				Vangerow}}, \bibinfo {author} {\bibfnamefont {F.}~\bibnamefont
			{Stienkemeier}}, \bibinfo {author} {\bibfnamefont {A.~S.}\ \bibnamefont
			{Bogomolov}}, \bibinfo {author} {\bibfnamefont {A.~V.}\ \bibnamefont
			{Baklanov}}, \bibinfo {author} {\bibfnamefont {D.~M.}\ \bibnamefont {Reich}},
		\bibinfo {author} {\bibfnamefont {W.}~\bibnamefont {Skomorowski}}, \bibinfo
		{author} {\bibfnamefont {C.~P.}\ \bibnamefont {Koch}}, \ and\ \bibinfo
		{author} {\bibfnamefont {M.}~\bibnamefont {Mudrich}},\ }\href {\doibase
		http://dx.doi.org/10.1063/1.4906512} {\bibfield  {journal} {\bibinfo
			{journal} {J. Chem. Phys.}\ }\textbf {\bibinfo {volume} {142}},\ \bibinfo
		{eid} {044303} (\bibinfo {year} {2015})}\BibitemShut {NoStop}%
	\bibitem [{\citenamefont {Jonah}(1971)}]{Jonah:1971}%
	\BibitemOpen
	\bibfield  {author} {\bibinfo {author} {\bibfnamefont {C.}~\bibnamefont
			{Jonah}},\ }\href@noop {} {\bibfield  {journal} {\bibinfo  {journal} {J.
				Chem. Phys.}\ }\textbf {\bibinfo {volume} {55}},\ \bibinfo {pages} {1915}
		(\bibinfo {year} {1971})}\BibitemShut {NoStop}%
	\bibitem [{\citenamefont {Yang}\ and\ \citenamefont
		{Bersohn}(1974)}]{Yang:1974}%
	\BibitemOpen
	\bibfield  {author} {\bibinfo {author} {\bibfnamefont {S.~C.}\ \bibnamefont
			{Yang}}\ and\ \bibinfo {author} {\bibfnamefont {R.}~\bibnamefont {Bersohn}},\
	}\href@noop {} {\bibfield  {journal} {\bibinfo  {journal} {J. Chem. Phys.}\
		}\textbf {\bibinfo {volume} {61}},\ \bibinfo {pages} {4400} (\bibinfo {year}
		{1974})}\BibitemShut {NoStop}%
	\bibitem [{\citenamefont {Peterka}\ \emph {et~al.}(2003)\citenamefont
		{Peterka}, \citenamefont {Lindinger}, \citenamefont {Poisson}, \citenamefont
		{Ahmed},\ and\ \citenamefont {Neumark}}]{Peterka:2003}%
	\BibitemOpen
	\bibfield  {author} {\bibinfo {author} {\bibfnamefont {D.~S.}\ \bibnamefont
			{Peterka}}, \bibinfo {author} {\bibfnamefont {A.}~\bibnamefont {Lindinger}},
		\bibinfo {author} {\bibfnamefont {L.}~\bibnamefont {Poisson}}, \bibinfo
		{author} {\bibfnamefont {M.}~\bibnamefont {Ahmed}}, \ and\ \bibinfo {author}
		{\bibfnamefont {D.~M.}\ \bibnamefont {Neumark}},\ }\href@noop {} {\bibfield
		{journal} {\bibinfo  {journal} {Phys. Rev. Lett.}\ }\textbf {\bibinfo
			{volume} {91}},\ \bibinfo {pages} {043401} (\bibinfo {year}
		{2003})}\BibitemShut {NoStop}%
	\bibitem [{\citenamefont {Buchta}\ \emph
		{et~al.}(2013{\natexlab{a}})\citenamefont {Buchta}, \citenamefont {Krishnan},
		\citenamefont {Brauer}, \citenamefont {Drabbels}, \citenamefont {O'Keeffe},
		\citenamefont {Devetta}, \citenamefont {Di~Fraia}, \citenamefont {Callegari},
		\citenamefont {Richter}, \citenamefont {Coreno}, \citenamefont {Prince},
		\citenamefont {Stienkemeier}, \citenamefont {Moshammer},\ and\ \citenamefont
		{Mudrich}}]{Buchta:2013}%
	\BibitemOpen
	\bibfield  {author} {\bibinfo {author} {\bibfnamefont {D.}~\bibnamefont
			{Buchta}}, \bibinfo {author} {\bibfnamefont {S.~R.}\ \bibnamefont
			{Krishnan}}, \bibinfo {author} {\bibfnamefont {N.~B.}\ \bibnamefont
			{Brauer}}, \bibinfo {author} {\bibfnamefont {M.}~\bibnamefont {Drabbels}},
		\bibinfo {author} {\bibfnamefont {P.}~\bibnamefont {O'Keeffe}}, \bibinfo
		{author} {\bibfnamefont {M.}~\bibnamefont {Devetta}}, \bibinfo {author}
		{\bibfnamefont {M.}~\bibnamefont {Di~Fraia}}, \bibinfo {author}
		{\bibfnamefont {C.}~\bibnamefont {Callegari}}, \bibinfo {author}
		{\bibfnamefont {R.}~\bibnamefont {Richter}}, \bibinfo {author} {\bibfnamefont
			{M.}~\bibnamefont {Coreno}}, \bibinfo {author} {\bibfnamefont {K.~C.}\
			\bibnamefont {Prince}}, \bibinfo {author} {\bibfnamefont {F.}~\bibnamefont
			{Stienkemeier}}, \bibinfo {author} {\bibfnamefont {R.}~\bibnamefont
			{Moshammer}}, \ and\ \bibinfo {author} {\bibfnamefont {M.}~\bibnamefont
			{Mudrich}},\ }\href {\doibase 10.1021/jp401424w} {\bibfield  {journal}
		{\bibinfo  {journal} {J. Phys. Chem. A}\ }\textbf {\bibinfo {volume} {117}},\
		\bibinfo {pages} {4394} (\bibinfo {year} {2013}{\natexlab{a}})}\BibitemShut
	{NoStop}%
	\bibitem [{\citenamefont {Buchta}\ \emph
		{et~al.}(2013{\natexlab{b}})\citenamefont {Buchta}, \citenamefont {Krishnan},
		\citenamefont {Brauer}, \citenamefont {Drabbels}, \citenamefont {O’Keeffe},
		\citenamefont {Devetta}, \citenamefont {Di~Fraia}, \citenamefont {Callegari},
		\citenamefont {Richter}, \citenamefont {Coreno}, \citenamefont {Prince},
		\citenamefont {Stienkemeier}, \citenamefont {Ullrich}, \citenamefont
		{Moshammer},\ and\ \citenamefont {Mudrich}}]{BuchtaJCP:2013}%
	\BibitemOpen
	\bibfield  {author} {\bibinfo {author} {\bibfnamefont {D.}~\bibnamefont
			{Buchta}}, \bibinfo {author} {\bibfnamefont {S.~R.}\ \bibnamefont
			{Krishnan}}, \bibinfo {author} {\bibfnamefont {N.~B.}\ \bibnamefont
			{Brauer}}, \bibinfo {author} {\bibfnamefont {M.}~\bibnamefont {Drabbels}},
		\bibinfo {author} {\bibfnamefont {P.}~\bibnamefont {O’Keeffe}}, \bibinfo
		{author} {\bibfnamefont {M.}~\bibnamefont {Devetta}}, \bibinfo {author}
		{\bibfnamefont {M.}~\bibnamefont {Di~Fraia}}, \bibinfo {author}
		{\bibfnamefont {C.}~\bibnamefont {Callegari}}, \bibinfo {author}
		{\bibfnamefont {R.}~\bibnamefont {Richter}}, \bibinfo {author} {\bibfnamefont
			{M.}~\bibnamefont {Coreno}}, \bibinfo {author} {\bibfnamefont {K.~C.}\
			\bibnamefont {Prince}}, \bibinfo {author} {\bibfnamefont {F.}~\bibnamefont
			{Stienkemeier}}, \bibinfo {author} {\bibfnamefont {J.}~\bibnamefont
			{Ullrich}}, \bibinfo {author} {\bibfnamefont {R.}~\bibnamefont {Moshammer}},
		\ and\ \bibinfo {author} {\bibfnamefont {M.}~\bibnamefont {Mudrich}},\ }\href
	{\doibase http://dx.doi.org/10.1063/1.4818531} {\bibfield  {journal}
		{\bibinfo  {journal} {J. Chem. Phys.}\ }\textbf {\bibinfo {volume} {139}},\
		\bibinfo {eid} {084301} (\bibinfo {year} {2013}{\natexlab{b}})}\BibitemShut
	{NoStop}%
	\bibitem [{\citenamefont {Rosenblit}\ and\ \citenamefont
		{Jortner}(2006)}]{Rosenblit:2006}%
	\BibitemOpen
	\bibfield  {author} {\bibinfo {author} {\bibfnamefont {M.}~\bibnamefont
			{Rosenblit}}\ and\ \bibinfo {author} {\bibfnamefont {J.}~\bibnamefont
			{Jortner}},\ }\href@noop {} {\bibfield  {journal} {\bibinfo  {journal} {J.
				Chem. Phys.}\ }\textbf {\bibinfo {volume} {124}},\ \bibinfo {pages} {194505}
		(\bibinfo {year} {2006})}\BibitemShut {NoStop}%
	\bibitem [{\citenamefont {Masson}\ \emph {et~al.}(2010)\citenamefont {Masson},
		\citenamefont {Poisson}, \citenamefont {Gaveau}, \citenamefont {Soep},
		\citenamefont {Mestdagh}, \citenamefont {Mazet},\ and\ \citenamefont
		{Spiegelman}}]{Masson:2010}%
	\BibitemOpen
	\bibfield  {author} {\bibinfo {author} {\bibfnamefont {A.}~\bibnamefont
			{Masson}}, \bibinfo {author} {\bibfnamefont {L.}~\bibnamefont {Poisson}},
		\bibinfo {author} {\bibfnamefont {M.-A.}\ \bibnamefont {Gaveau}}, \bibinfo
		{author} {\bibfnamefont {B.}~\bibnamefont {Soep}}, \bibinfo {author}
		{\bibfnamefont {J.-M.}\ \bibnamefont {Mestdagh}}, \bibinfo {author}
		{\bibfnamefont {V.}~\bibnamefont {Mazet}}, \ and\ \bibinfo {author}
		{\bibfnamefont {F.}~\bibnamefont {Spiegelman}},\ }\href@noop {} {\bibfield
		{journal} {\bibinfo  {journal} {J. Chem. Phys.}\ }\textbf {\bibinfo {volume}
			{133}},\ \bibinfo {pages} {054307} (\bibinfo {year} {2010})}\BibitemShut
	{NoStop}%
	\bibitem [{\citenamefont {Masson}\ \emph {et~al.}(2014)\citenamefont {Masson},
		\citenamefont {Heitz}, \citenamefont {Mestdagh}, \citenamefont {Gaveau},
		\citenamefont {Poisson},\ and\ \citenamefont {Spiegelman}}]{Masson:2014}%
	\BibitemOpen
	\bibfield  {author} {\bibinfo {author} {\bibfnamefont {A.}~\bibnamefont
			{Masson}}, \bibinfo {author} {\bibfnamefont {M.-C.}\ \bibnamefont {Heitz}},
		\bibinfo {author} {\bibfnamefont {J.-M.}\ \bibnamefont {Mestdagh}}, \bibinfo
		{author} {\bibfnamefont {M.-A.}\ \bibnamefont {Gaveau}}, \bibinfo {author}
		{\bibfnamefont {L.}~\bibnamefont {Poisson}}, \ and\ \bibinfo {author}
		{\bibfnamefont {F.}~\bibnamefont {Spiegelman}},\ }\href@noop {} {\bibfield
		{journal} {\bibinfo  {journal} {Phys. Rev. Lett.}\ }\textbf {\bibinfo
			{volume} {113}},\ \bibinfo {pages} {123005} (\bibinfo {year}
		{2014})}\BibitemShut {NoStop}%
\end{thebibliography}
\end{document}